\documentclass[12pt,a4paper]{article}

\usepackage{ifthen} 
\newboolean{pdflatex}
\setboolean{pdflatex}{true} 

\newboolean{articletitles}
\setboolean{articletitles}{true} 

\newboolean{uprightparticles}
\setboolean{uprightparticles}{false} 

\usepackage{lineno}  
\usepackage{xspace} 

\usepackage{subfigure}
\usepackage{graphicx}  
\usepackage[usenames,dvipsnames]{color}
\usepackage{colortbl}
\graphicspath{{./figs/}} 
\usepackage{dcolumn}
\usepackage{amsmath} 
\usepackage{amssymb}
\usepackage{amsfonts}
\usepackage{upgreek} 
\usepackage{epstopdf}

\newcommand*\patchAmsMathEnvironmentForLineno[1]{%
\expandafter\let\csname old#1\expandafter\endcsname\csname #1\endcsname
\expandafter\let\csname oldend#1\expandafter\endcsname\csname
end#1\endcsname
 \renewenvironment{#1}%
   {\linenomath\csname old#1\endcsname}%
   {\csname oldend#1\endcsname\endlinenomath}%
}
\newcommand*\patchBothAmsMathEnvironmentsForLineno[1]{%
  \patchAmsMathEnvironmentForLineno{#1}%
  \patchAmsMathEnvironmentForLineno{#1*}%
}
\AtBeginDocument{%
\patchBothAmsMathEnvironmentsForLineno{equation}%
\patchBothAmsMathEnvironmentsForLineno{align}%
\patchBothAmsMathEnvironmentsForLineno{flalign}%
\patchBothAmsMathEnvironmentsForLineno{alignat}%
\patchBothAmsMathEnvironmentsForLineno{gather}%
\patchBothAmsMathEnvironmentsForLineno{multline}%
}

\usepackage{hyperref}    
\usepackage[all]{hypcap} 

\usepackage{lhcb-symbols-def} 

\usepackage{cite} 
\usepackage{mciteplus}

\begin{document}

\renewcommand{\thefootnote}{\fnsymbol{footnote}}
\setcounter{footnote}{1}


\begin{titlepage}
\pagenumbering{roman}

\centerline{\large EUROPEAN ORGANIZATION FOR NUCLEAR RESEARCH (CERN)}
\vspace*{1.5cm}
\hspace*{-0.5cm}
\begin{tabular*}{\linewidth}{lc@{\extracolsep{\fill}}r}
\ifthenelse{\boolean{pdflatex}}
{\vspace*{-2.7cm}\mbox{\!\!\!\includegraphics[width=.14\textwidth]{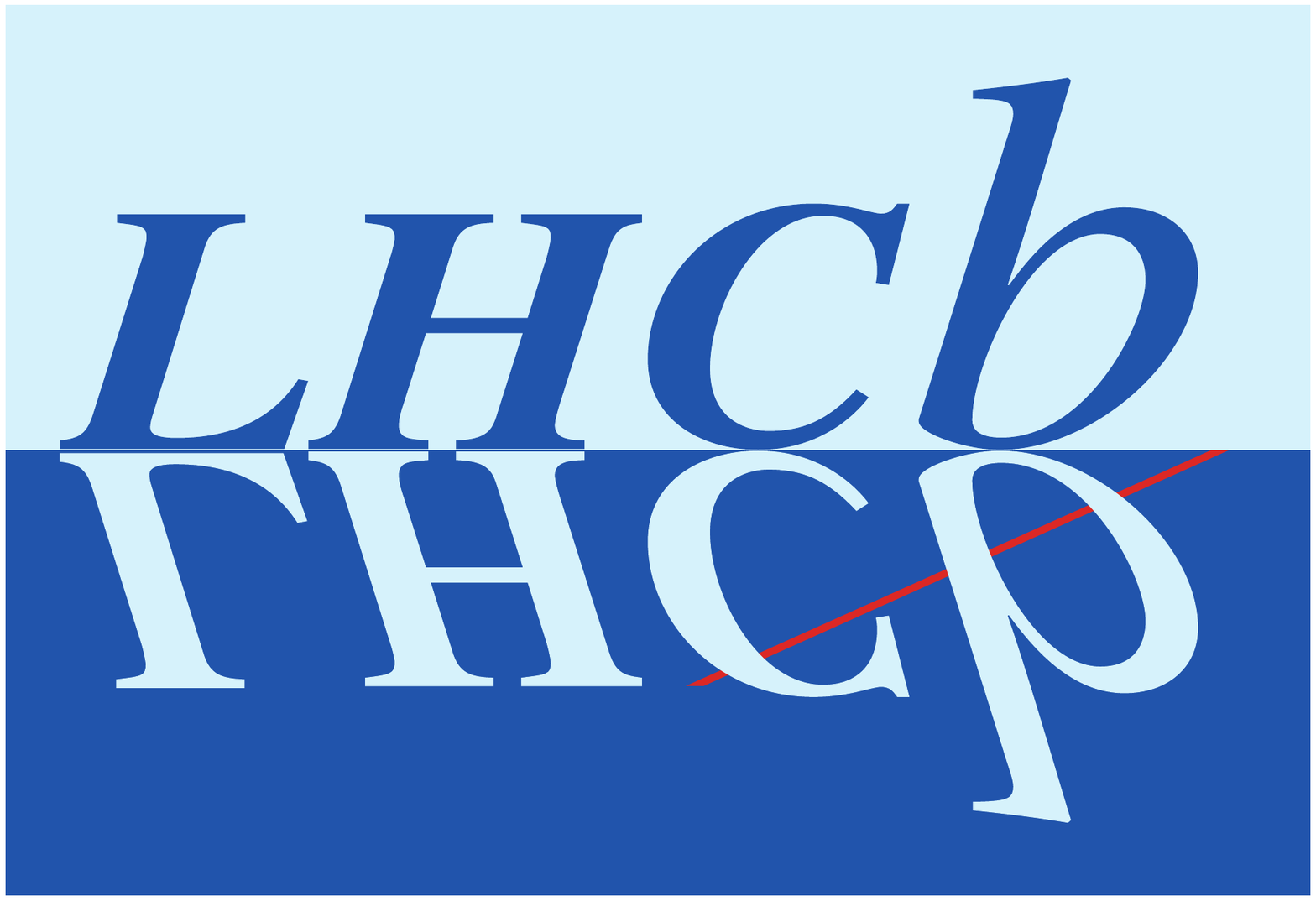}} & &}%
{\vspace*{-1.2cm}\mbox{\!\!\!\includegraphics[width=.12\textwidth]{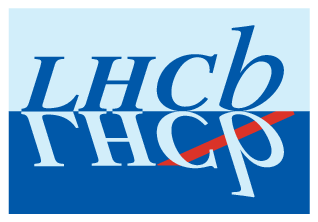}} & &}%
\\
 & & CERN-PH-EP-2012-268 \\  
 & & LHCb-PAPER-2012-027 \\  
& & 26 September 2012 \\ 
& & \\
\end{tabular*}

\vspace*{1.5cm}

{\bf\boldmath\huge
\begin{center}
A model-independent Dalitz \\ plot analysis of $B^\pm \to D K^\pm$ with $D \to \KS h^+h^-$ ($h=\pi, K$) decays and constraints on the CKM angle $\gamma$
\end{center}
}

\vspace*{0.5cm}

\begin{center}
The LHCb collaboration\footnote{Authors are listed on the following pages.}
\end{center}

\vspace{\fill}

\vspace{-0.25cm}
\begin{abstract}
  \noindent
A binned Dalitz plot analysis of $B^\pm \to D K^\pm$ decays, with $D \to \KS \pi^+\pi^-$ and $D \to \KS  K^+ K^-$, is performed to measure the \CP-violating observables $x_{\pm}$ and $y_{\pm}$ which are sensitive to  the CKM angle $\gamma$.  The analysis exploits 1.0~$\rm fb^{-1}$ of data collected by the LHCb experiment.  The study makes no model-based assumption on the variation of the strong phase of the $D$ decay amplitude over the Dalitz plot, but uses measurements of this quantity from CLEO-c as input.
The values of the parameters are found to be $x_- =   (0.0 \pm 4.3 \pm 1.5 \pm 0.6) \times 10^{-2}$,  $y_-  = (2.7 \pm 5.2 \pm 0.8 \pm 2.3) \times 10^{-2}$, $x_+   =   ( -10.3 \pm 4.5 \pm 1.8 \pm 1.4 )\times 10^{-2}$ and  $y_+  = (-0.9 \pm 3.7 \pm 0.8 \pm 3.0)\times 10^{-2}$. The first, second, and third uncertainties are the statistical, the experimental systematic, and the error associated with the precision of the strong-phase parameters measured at CLEO-c, respectively.
These results correspond to $\gamma = (44^{\,+43}_{\,-38})^\circ$, with a second solution at $\gamma \to \gamma + 180^\circ$, and $r_B = 0.07 \pm 0.04$, where $r_B$ is the ratio between the suppressed and favoured $B$ decay amplitudes.    
\end{abstract}

\vspace*{0.1cm}

\begin{center}
  Submitted to Physics Letters B
\end{center}

\vspace{\fill}

\end{titlepage}


\newpage
\setcounter{page}{2}
\mbox{~}
\newpage

\centerline{\large\bf LHCb collaboration}
\begin{flushleft}
\small
R.~Aaij$^{38}$, 
C.~Abellan~Beteta$^{33,n}$, 
A.~Adametz$^{11}$, 
B.~Adeva$^{34}$, 
M.~Adinolfi$^{43}$, 
C.~Adrover$^{6}$, 
A.~Affolder$^{49}$, 
Z.~Ajaltouni$^{5}$, 
J.~Albrecht$^{35}$, 
F.~Alessio$^{35}$
% ...

{\footnotesize \it
% ...
$ ^{5}$Clermont Universit\'{e}, Universit\'{e} Blaise Pascal, CNRS/IN2P3, LPC, Clermont-Ferrand, France\\
$ ^{6}$CPPM, Aix-Marseille Universit\'{e}, CNRS/IN2P3, Marseille, France\\
% ...
$ ^{11}$Physikalisches Institut, Ruprecht-Karls-Universit\"{a}t Heidelberg, Heidelberg, Germany\\
% ...
$ ^{33}$Universitat de Barcelona, Barcelona, Spain\\
$ ^{34}$Universidad de Santiago de Compostela, Santiago de Compostela, Spain\\
$ ^{35}$European Organization for Nuclear Research (CERN), Geneva, Switzerland\\
% ...
$ ^{38}$Nikhef National Institute for Subatomic Physics, Amsterdam, The Netherlands\\
% ...
$ ^{43}$H.H. Wills Physics Laboratory, University of Bristol, Bristol, United Kingdom\\
% ...
$ ^{49}$Oliver Lodge Laboratory, University of Liverpool, Liverpool, United Kingdom\\
% ...
$ ^{n}$LIFAELS, La Salle, Universitat Ramon Llull, Barcelona, Spain\\
% ...
}
\end{flushleft}

\cleardoublepage


\renewcommand{\thefootnote}{\arabic{footnote}}
\setcounter{footnote}{0}



\pagestyle{plain} 
\setcounter{page}{1}
\pagenumbering{arabic}


%

\section{Introduction}
\label{sec:Introduction}
A precise determination of the Unitarity Triangle angle $\gamma$  (also denoted as $\phi_3$), is an important goal in flavour physics. Measurements of this weak phase in tree-level processes involving the interference between  $b \to c \bar{u}s$ and $b \to u \bar{c} s$ transitions are expected to be insensitive to new physics contributions, thereby providing a Standard Model benchmark against which other observables, more likely to be affected by new physics, can be compared.  A powerful approach for measuring $\gamma$ is to study
 \CP-violating observables in $B^\pm \to D K^\pm$ decays, where $D$ designates a neutral $D$ meson reconstructed in a final state 
common to both $D^0$ and \Dzb decays.  Examples of such final states include two-body modes, where \lhcb has already presented results~\cite{LHCBADS}, and self \CP-conjugate three-body decays, such as $\KS \pi^+\pi^-$ and $\KS K^+K^-$, designated collectively as $\KS h^+h^-$.

The proposal to measure $\gamma$ with $B^\pm \to D K^\pm$, $D \to \KS h^+h^-$ decays was first made in Refs.~\cite{GGSZ,BONDARGGSZ}.  The strategy relies on comparing the distribution of events in the $D \to \KS h^+h^-$ Dalitz plot
for $B^+ \to D K^+$ and $B^- \to D K^-$ decays.  However, in order to determine $\gamma$ it is necessary to know how the strong phase of the $D$ decay varies over the Dalitz plot.  One approach for solving this problem, adopted by \babar~\cite{BABAR2005,BABAR2008,BABAR2010} and Belle~\cite{BELLE2004,BELLE2006,BELLE2010}, is to use an amplitude model fitted on flavour-tagged $D \to \KS h^+h^-$ decays to provide this input.  An attractive alternative~\cite{GGSZ,BPMODIND1,BPMODIND2} is to make use of direct measurements of the strong phase behaviour in bins of the Dalitz plot, which can be obtained from quantum-correlated $D\Db$ pairs from $\psi(3770)$ decays and that are available from CLEO-c~\cite{CLEOCISI}, thereby avoiding the need to assign any model-related systematic uncertainty.  A first model-independent analysis was recently presented by Belle~\cite{BELLEMODIND} using $B^\pm \to D K^\pm$, $D \to \KS \pi^+\pi^-$ decays.
In this Letter, $pp$ collision data at $\sqrt{s}=7$~\tev, corresponding to an integrated luminosity of $1.0~{\rm fb^{-1}}$   and accumulated by \lhcb in 2011, are exploited to perform a similar model-independent study of the decay mode $B^\pm \to D K^\pm$ with $D \to \KS \pi^+\pi^-$ and $D \to \KS K^+K^-$.  The results are used to set constraints on the value of $\gamma$.


\section{Formalism and external inputs}
\label{sec:formalism}

The amplitude of the decay $B^+ \to D K^+$, $D \to \KS h^+h^-$ can be written as the superposition of the $B^+ \to \Dzb K^+$ and  $B^+ \to D^0 K^+$ contributions as
\begin{equation}
A_B (m_+^2, m_-^2) = \overline{A} + r_B e^{i(\delta_B + \gamma)}A.
\label{eq:bamplitude}
\end{equation}
Here  $m_+^2$ and $m_-^2$ are the  invariant masses squared of the $\KS h^+$ and  $\KS h^-$ combinations, respectively,
that define the position of the decay in the Dalitz plot, $A=A(m_+^2,m_-^2)$ is the $D^0 \to \KS h^+ h^-$ amplitude, 
 and $\overline{A}=\overline{A}(m_+^2,m_-^2)$ the  $\Dzb \to \KS h^+ h^-$ amplitude.
 The parameter $r_B$, the ratio of the magnitudes of the $B^+ \to D^0 K^+$ and $B^+ \to \Dzb K^+$ amplitudes, is $\sim$0.1~\cite{HFAG}, and $\delta_B$ is the strong-phase difference between them.  The equivalent expression for the charge-conjugated decay $B^- \to D K^-$ is obtained by making the substitutions $\gamma \to -\gamma$ and $A \leftrightarrow \overline{A}$.  Neglecting \CP violation, which is known to be small in $D^0 - \Dzb$ mixing and Cabibbo-favoured $D$ meson decays~\cite{PDG}, the conjugate amplitudes are related by $A(m_+^2,m_-^2) = \overline{A}(m_-^2,m_+^2)$.

Following the formalism set out in Ref.~\cite{GGSZ}, the Dalitz plot is partitioned into $2N$ regions symmetric under the exchange $m_+^2 \leftrightarrow m_-^2$.  The bins are labelled from $-N$ to $+N$ (excluding zero), where the positive bins satisfy $m_-^2 > m_+^2$.  At each point in the Dalitz plot, there is a strong-phase difference $\delta_D(m_+^2,m_-^2) = \arg \overline{A} - \arg {A}$ between the \Dzb and $\Dz$ decay.  The cosine of the strong-phase difference averaged in each bin and weighted by the absolute decay rate is termed $c_i$ and is given by
\begin{equation}
c_i = \frac{\int_{{\cal D}_i} (|A||\overline{A}| \cos{\delta_D})\, d{\cal D}}
{\sqrt{\int_{{\cal D}_i} |A|^2 \,d{\cal D}} \,\sqrt{ \int_{{\cal D}_i} |\overline{A}|^2 \,d{\cal D}}},
\end{equation}
where the integrals are evaluated over the area ${\cal D}$ of bin $i$.  An analogous expression may be written for $s_i$, which is the 
sine of the strong-phase difference within bin $i$, weighted by the decay rate.  
The values of $c_i$ and $s_i$ can be determined by assuming a functional form for $|A|$, $|\overline{A}|$ and $\delta_D$, which may be obtained from an amplitude model fitted to flavour-tagged $D^0$ decays.  Alternatively direct measurements of $c_i$ and $s_i$ can be used. Such measurements have been performed at CLEO-c, exploiting quantum-correlated $D\Db$ pairs produced at the $\psi(3770)$ resonance. This has been done with a double-tagged method in which one $D$ meson is reconstructed in a decay to either $\KS h^+ h^-$ or $\KL h^+ h^-$, and the other $D$ meson is reconstructed either in a \CP eigenstate or in a decay to $ \KS h^+ h^-$. The efficiency-corrected event yields, combined with flavour-tag information, allow $c_i$ and $s_i$ to be determined~\cite{GGSZ,BPMODIND1,BPMODIND2}.
The latter approach is attractive as it avoids any assumption about the nature of the intermediate resonances which contribute to the $\KS h^+h^-$ final state; such an assumption leads to a systematic uncertainty associated with the variation in $\delta_D$ that is difficult to quantify.  Instead, an uncertainty is assigned that is related to the precision of the $c_i$ and $s_i$ measurements.

The population of each positive (negative) bin in the Dalitz plot arising from $B^+$ decays is $N_{+i}^+$ ($N_{-i}^+$), and that from $B^-$ decays is $N_{+i}^-$ ($N_{-i}^-$).  From Eq.~(\ref{eq:bamplitude}) it follows that
\begin{eqnarray}
N_{\pm i}^+ &=& h_{B^+} \left[ K_{\mp i} + (x_+^2 +y_+^2)K_{\pm i} + 2 \sqrt{K_i K_{-i}} ( x_+ c_{ \pm i} \mp y_+ s_{\pm i}) \right], \nonumber \\
N_{\pm i}^- &=& h_{B^-} \left[ K_{\pm i} + (x_-^2 + y_-^2) K_{\mp i} + 2 \sqrt{K_i K_{-i}} ( x_- c_{\pm i} \pm y_- s_{\pm i}) \right],
\label{eq:populations}
\end{eqnarray}
where $h_{B^\pm}$ are normalisation factors which can, in principle, be different for $B^+$ and $B^-$ due to the production asymmetries, and $K_i$ is the number of events in bin $i$ of the decay of a flavour tagged $D^0 \to \KS h^+ h^-$ Dalitz plot.  The sensitivity to $\gamma$ enters through the Cartesian parameters
\begin{equation}
x_\pm = r_B \cos (\delta_B \pm \gamma) { \rm\ and\  } \; y_\pm = r_B \sin (\delta_B \pm \gamma).
\label{eq:xydefinitions}
\end{equation}

In this analysis the observed distribution of candidates over the $D \to \KS h^+ h^-$ Dalitz plot is used to fit $x_\pm$, $y_\pm$ and  $h_{B^\pm}$.  The parameters $c_i$ and $s_i$ are taken from measurements performed by CLEO-c~\cite{CLEOCISI}.  
In this manner the analysis avoids any dependence on an amplitude model to describe the variation of the strong phase over the Dalitz plot.  
A model is used, however, to provide the input values for $K_i$. For the $\Dz\to\KS\pip\pim$ decay the model is taken from Ref.~\cite{BABAR2008} and for the $\Dz\to\KS\Kp\Km$ decay the model is taken from Ref.~\cite{BABAR2010}. This choice incurs no significant systematic uncertainty as the models have been shown to describe well the  intensity distribution of flavour-tagged $D^0$ decay data. 

The effect of $D^0 -\Dzb$ mixing is ignored in the above discussion, and was neglected in the CLEO-c measurements of $c_i$ and $s_i$ as well as in the construction of the amplitude model used to calculate $K_i$.  This leads to a bias of the order of $0.2^\circ$ in the $\gamma$ determination~\cite{BPV} which is negligible for the current analysis.

The CLEO-c study segments the $\KS \pi^+\pi^-$ Dalitz plot into $2 \times 8$ bins. Several bin definitions are available.  
Here the `optimal binning' variant is adopted.  In this scheme the bins have been chosen to optimise the statistical sensitivity
to $\gamma$ in the presence of a low level of background, which is appropriate for this analysis.
The optimisation has been performed assuming a strong-phase difference distribution as predicted by the \babar model presented in Ref.~\cite{BABAR2008}.
The use of a specific model in defining the bin boundaries does not bias the $c_i$ and $s_i$ measurements.  If the model is a poor description of the underlying decay the only consequence will be to reduce the statistical sensitivity of the $\gamma$ measurement.

\begin{figure}[htbp]
\begin{center}
\includegraphics[width=0.48\textwidth]{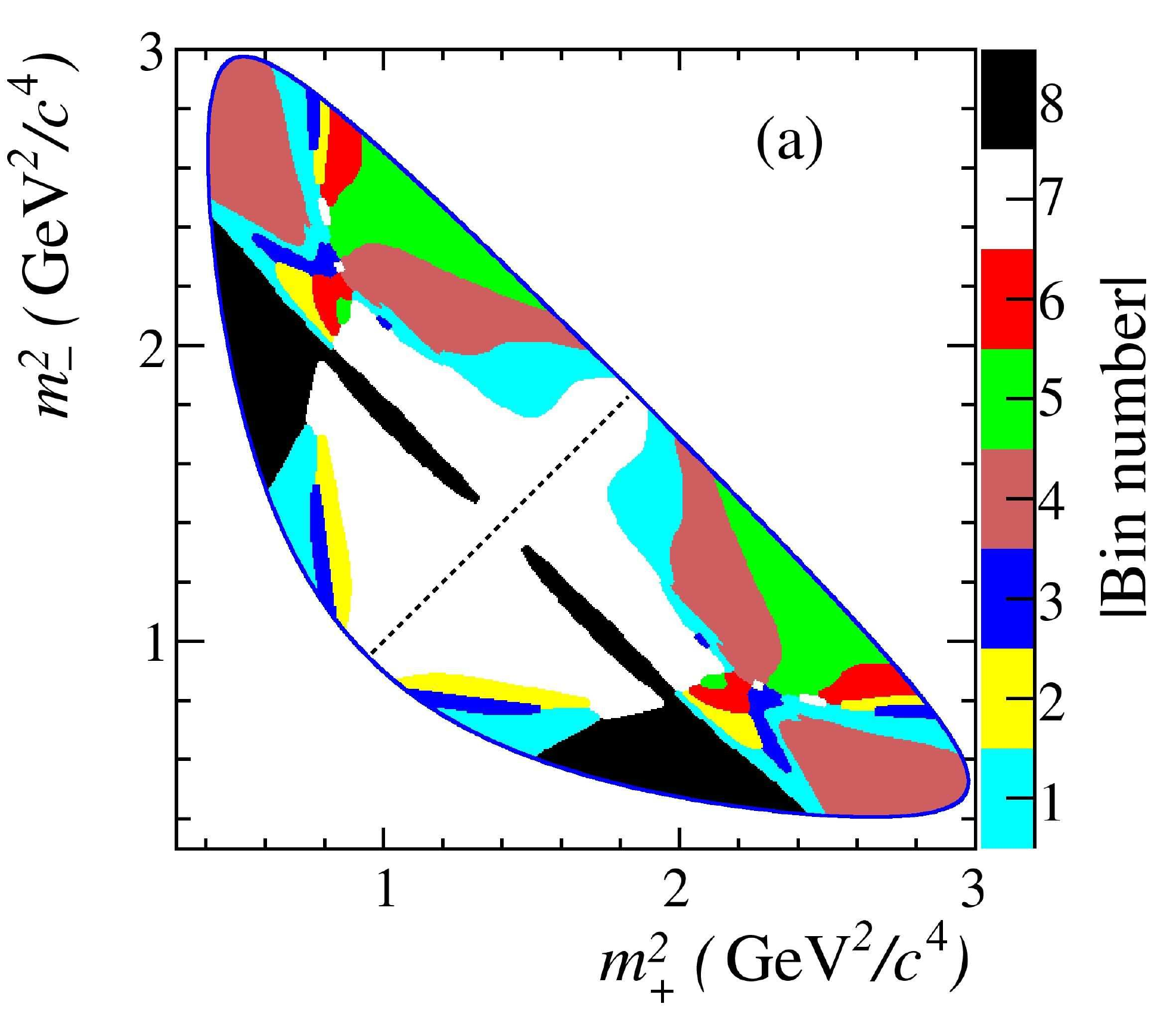}
\includegraphics[width=0.48\textwidth]{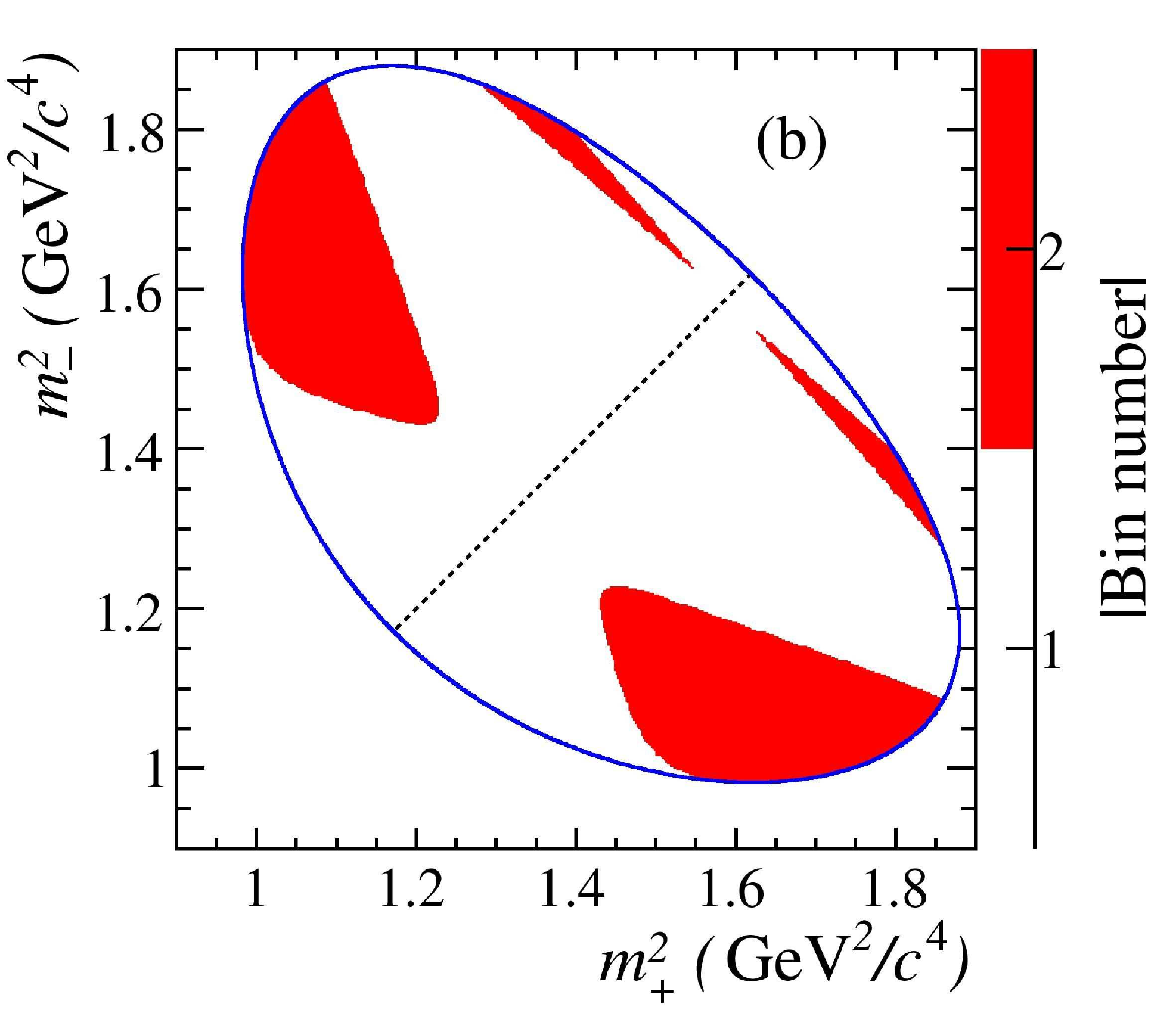}
\caption{\small Binning choices for (a) $D \to \KS \pi^+\pi^-$ and (b) $D \to \KS K^+K^-$. The diagonal line separates the positive and negative bins.}
\label{fig:bins}
\end{center}
\end{figure}

 For the \KSKK final state $c_i$ and $s_i$ measurements are available for the Dalitz plot partitioned into  $2 \times 2$,  $2 \times 3$ and  $2 \times 4$ bins,  with the guiding model being that from the \babar study described in  Ref.~\cite{BABAR2010}. The bin boundaries divide the Dalitz plot into bins of equal size with respect to the strong-phase difference between the $D^0$ and \Dzb amplitudes.
The current analysis adopts the $2 \times 2$ option, a decision driven by the size of the signal sample. 
The binning choices for the two decay modes are shown in Fig.~\ref{fig:bins}.

\section{The \lhcb detector}
\label{sec:detector}

The \lhcb detector~\cite{Alves:2008zz} is a single-arm forward
spectrometer covering the \mbox{pseudorapidity} range $2<\eta <5$. The
detector includes a high precision tracking system consisting of a
silicon-strip vertex detector surrounding the $pp$ interaction region,
a large-area silicon-strip detector (VELO) located upstream of a dipole
magnet with a bending power of about $4{\rm\,Tm}$, and three stations
of silicon-strip detectors and straw drift-tubes placed
downstream. The combined tracking system has a momentum resolution
of (0.4 -- 0.6)\% in the range of 5 -- 100~\gevc,
and an impact parameter (IP) resolution of 20~\mum for tracks with high
transverse momentum (\pt).
The dipole magnet can
be operated in either polarity and this feature is
used to reduce systematic effects due to detector
asymmetries. In the data set considered in this analysis, 58\% of data were taken
with one polarity and 42\% with the other.
Charged hadrons are identified using two
ring-imaging Cherenkov (RICH) detectors. Photon, electron and hadron
candidates are identified by a calorimeter system consisting of
scintillating-pad and preshower detectors, an electromagnetic
calorimeter and a hadronic calorimeter. Muons are identified by a system composed of alternating layers of iron and multiwire
proportional chambers. 

A two-stage trigger is employed. First a
hardware-based decision is taken at a frequency
up to 40 MHz. It accepts high transverse energy
clusters in either the electromagnetic calorimeter or
hadron calorimeter, or a muon of high \pt. For this analysis, it is required
that one of the charged final-state tracks forming the $B^\pm$ candidate
points at a deposit in the hadron calorimeter,
or that the hardware-trigger decision was taken
independently of these tracks. A second trigger
level, implemented in software, receives
1~MHz of events and retains $\sim$0.3\% of them~\cite{trigger}. It
searches for a track with large \pt and large IP with respect to any $pp$ interaction point which is called a primary vertex (PV). This
track is then required to be part of a two-, three- or four-track secondary
vertex with a high \pt sum, significantly displaced
from any PV. In order to maximise efficiency at
an acceptable trigger rate, the displaced vertex
is selected with a decision tree algorithm that
uses \pt, impact parameter, flight distance and track separation
information. Full event reconstruction occurs
offline, and a loose preselection is applied.

Approximately three million simulated events for each of the modes
$B^\pm \to D(\KS \pi^+\pi^-)K^\pm$ and 
$B^\pm \to D(\KS \pi^+\pi^-)\pi^\pm$ ,
and one million simulated events for each of 
$B^\pm \to D(\KS K^+ K^-)K^\pm$ and 
$B^\pm \to D(\KS K^+ K^-)\pi^\pm$ are used in 
the analysis,  as
well as a large inclusive sample of generic $B \to DX$ decays for background studies. 
These samples are generated using a version of \pythia~6.4~\cite{Sjostrand:2006za} tuned to model the $pp$ collisions~\cite{LHCb-PROC-2010-056}.
 \evtgen~\cite{Lange:2001uf} encodes the particle decays in which final state
radiation is generated using \photos~\cite{Golonka:2005pn}.
The interaction of the generated particles with the detector and its
response are implemented using the \geant
toolkit~\cite{Allison:2006ve, *Agostinelli:2002hh} as described in
Ref.~\cite{LHCb-PROC-2011-006}.

\section{Event selection and invariant mass spectrum fit}
\label{sec:selection}

Selection requirements are applied to isolate both $B^\pm \to D K^\pm$ and $B^\pm \to D \pi^\pm$ candidates, with $D \to \KS h^+h^-$.  Candidates selected in the Cabibbo-favoured  $B^\pm \to D \pi^\pm$ decay mode provide an important control sample which is exploited in the analysis.  

A production vertex is assigned to each \B candidate. This is the PV for which the reconstructed \B trajectory has the smallest IP $\chi^2$, where this quantity is defined as the difference in the $\chi^2$ fit of the PV with and without the tracks of the considered particle. The \KS candidates are formed from two oppositely charged tracks reconstructed in the tracking stations, either with associated hits in the VELO detector (long \KS candidate) or without (downstream \KS candidate).  The IP $\chi^2$ with respect to the PV of each of the long (downstream) \KS daughters is required to be greater than 16 (4). The angle $\theta$ between the \KS candidate momentum and the vector between the decay vertex and the PV, expected to be small given the high momentum of the $B$ meson, is required to satisfy $\cos \theta>0.99$, reducing background from combinations of random tracks. 

The $D$ meson candidates are reconstructed by combining the long (downstream) \KS candidates with two oppositely charged tracks for which the values of the  IP $\chi^2$ with respect to the PV are greater than 9 (16).  In the case of the $D \to \KS K^+K^-$ a loose particle identification (PID) requirement is placed on the kaons to reduce combinatoric backgrounds.
The IP $\chi^2$ of the candidate $D$ with respect to any PV is demanded to be greater than 9 in order to suppress directly produced $D$ mesons, and the angle $\theta$ between the $D$ candidate momentum and the vector between the decay and PV is required to satisfy the same criterion as for the \KS selection ($\cos\theta > 0.99$). The invariant mass resolution of the signal is $8.7$~\mevcc ($11.9$~\mevcc) for $D$ mesons reconstructed with long (downstream) \KS candidates, and a common window of $\pm 25$~\mevcc is imposed around the world average $D^0$ mass~\cite{PDG}. The \KS mass is determined after the addition of a constraint that the invariant mass of the two $D$ daughter pions or kaons and the two \KS daughter pions have the world average $D$ mass. The invariant mass resolution is $2.9$~\mevcc ($4.8$~\mevcc) for long (downstream) \KS decays. Candidates are retained for which the invariant mass of the two \KS daughters lies within $\pm 15$~\mevcc  of the world average \KS mass~\cite{PDG}.

The $D$ meson is combined with a candidate kaon or pion bachelor particle to form the \B candidate. The IP $\chi^2$ of the bachelor with respect to the PV  is required to be greater than 25. In order to ensure good  discrimination between pions and kaons in the RICH system only tracks with momentum less than $100$~\gevc are considered.  The bachelor is considered as a candidate kaon (pion) according to whether it passes (fails) a cut placed on the output of the RICH PID algorithm. The PID information is quantified as a difference between the logarithm of the likelihood under the mass hypothesis of a pion or a kaon.  Criteria are then imposed on the \B candidate: that the angle between its momentum and the vector between the decay and the PV should have a cosine greater than 0.9999 for candidates containing long \KS decays (0.99995 for downstream \KS decays); that the \B vertex-separation $\chi^2$ with respect to its PV is greater than 169; and that the \B IP $\chi^2$ with respect to the PV is less than 9.  To suppress background from charmless $B$ decays it is required that the $D$ vertex lies downstream of the $B$ vertex. In the events with a long \KS candidate, a further background arises from $B^\pm \to Dh^\pm$, $D\to \pi^+\pi^-h^+h^-$ decays, where the two pions are reconstructed as a long \KS candidate. This background is removed by requiring that the flight significance between the $D$ and \KS vertices is greater than 10. 

In order to obtain the best possible resolution in the Dalitz plot of the $D$ decay, and to provide further background suppression, the \B, $D$ and \KS vertices are refitted with additional constraints on the $D$ and \KS masses, and the \B momentum is required to point back to the PV.  The $\chi^2$ per degree of freedom of the fit is required to be less than 5.

Less than 0.4\% of the selected events are found to contain two or more candidates.  In these events only the $B$ candidate with the lowest
$\chi^2$ per degree of freedom from the refit is retained for subsequent study. In addition, 0.4\% of the candidates are found to have been reconstructed such that their $D$ Dalitz plot coordinates lie outside the defined bins, and these too are discarded.

\begin{figure}[htb]
\centering
\includegraphics[width=0.48\textwidth]{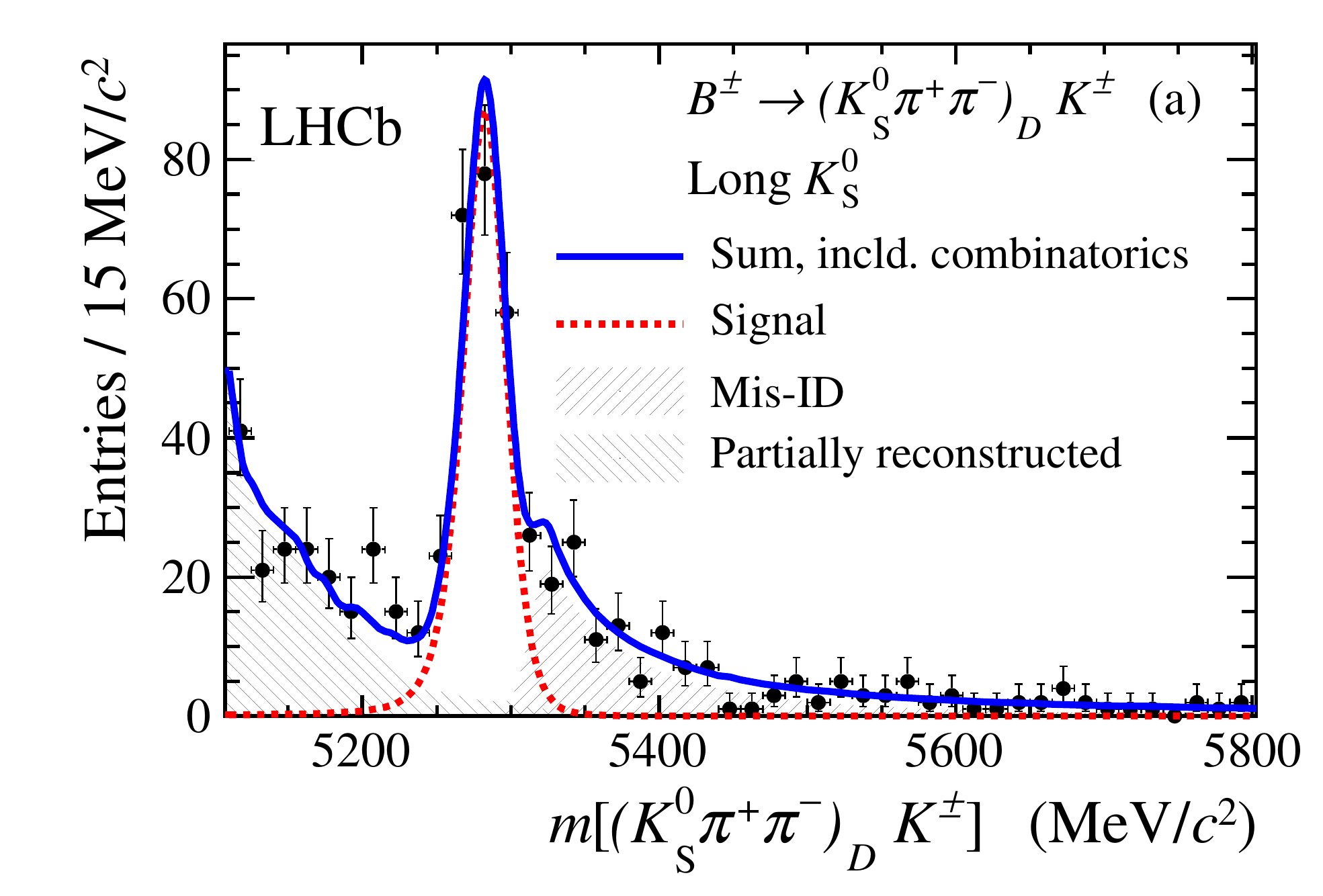}
\includegraphics[width=0.48\textwidth]{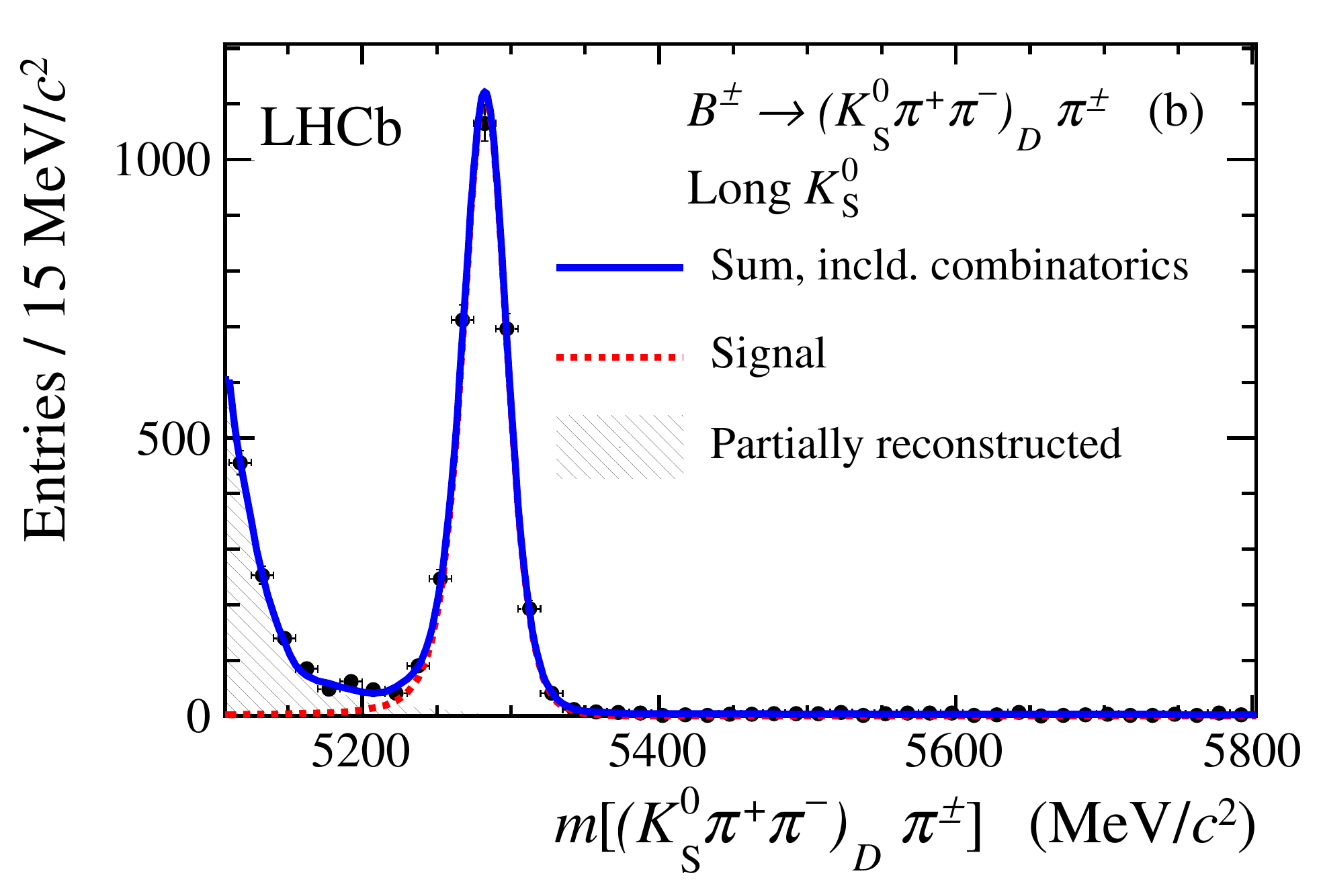}
\includegraphics[width=0.48\textwidth]{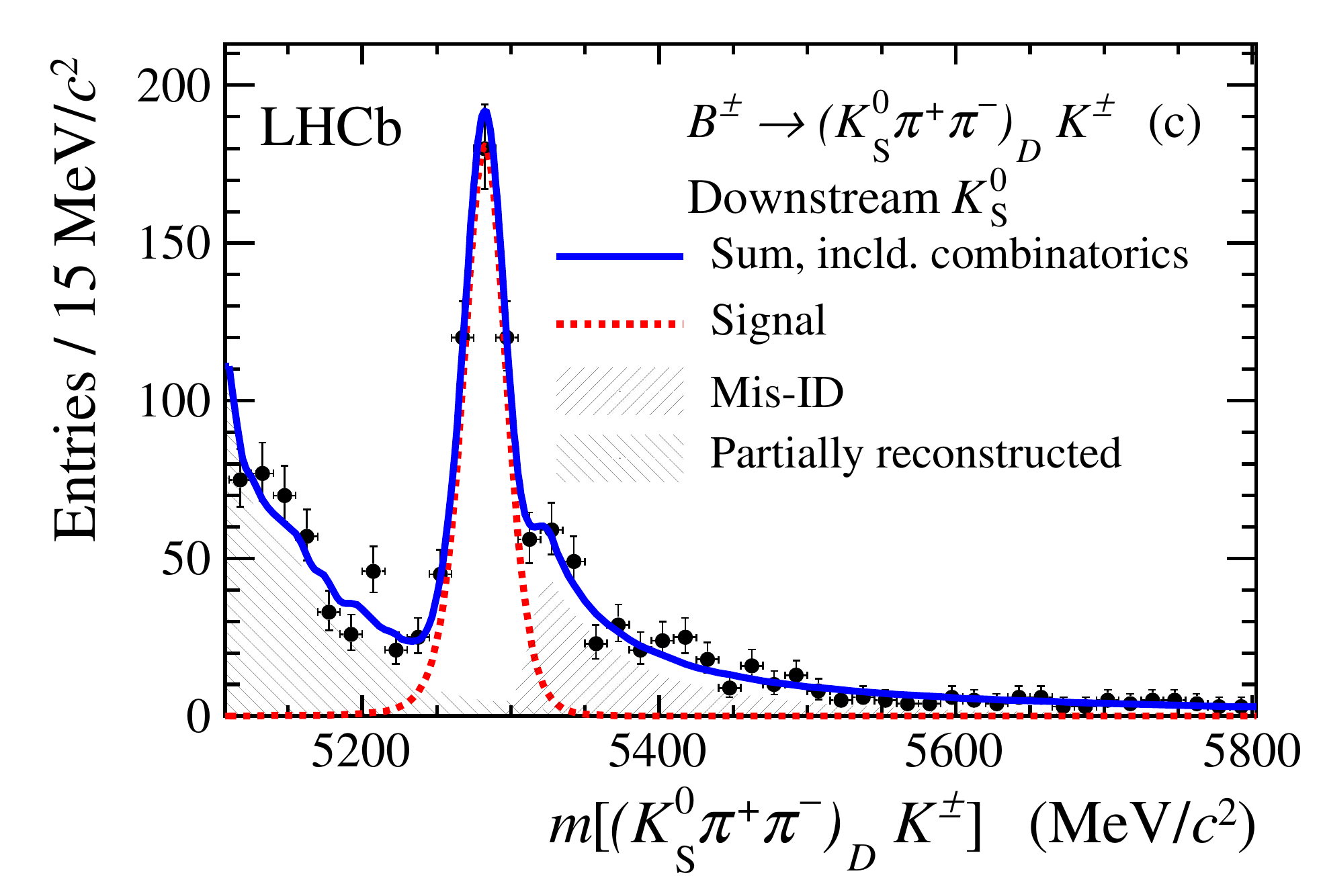}
\includegraphics[width=0.48\textwidth]{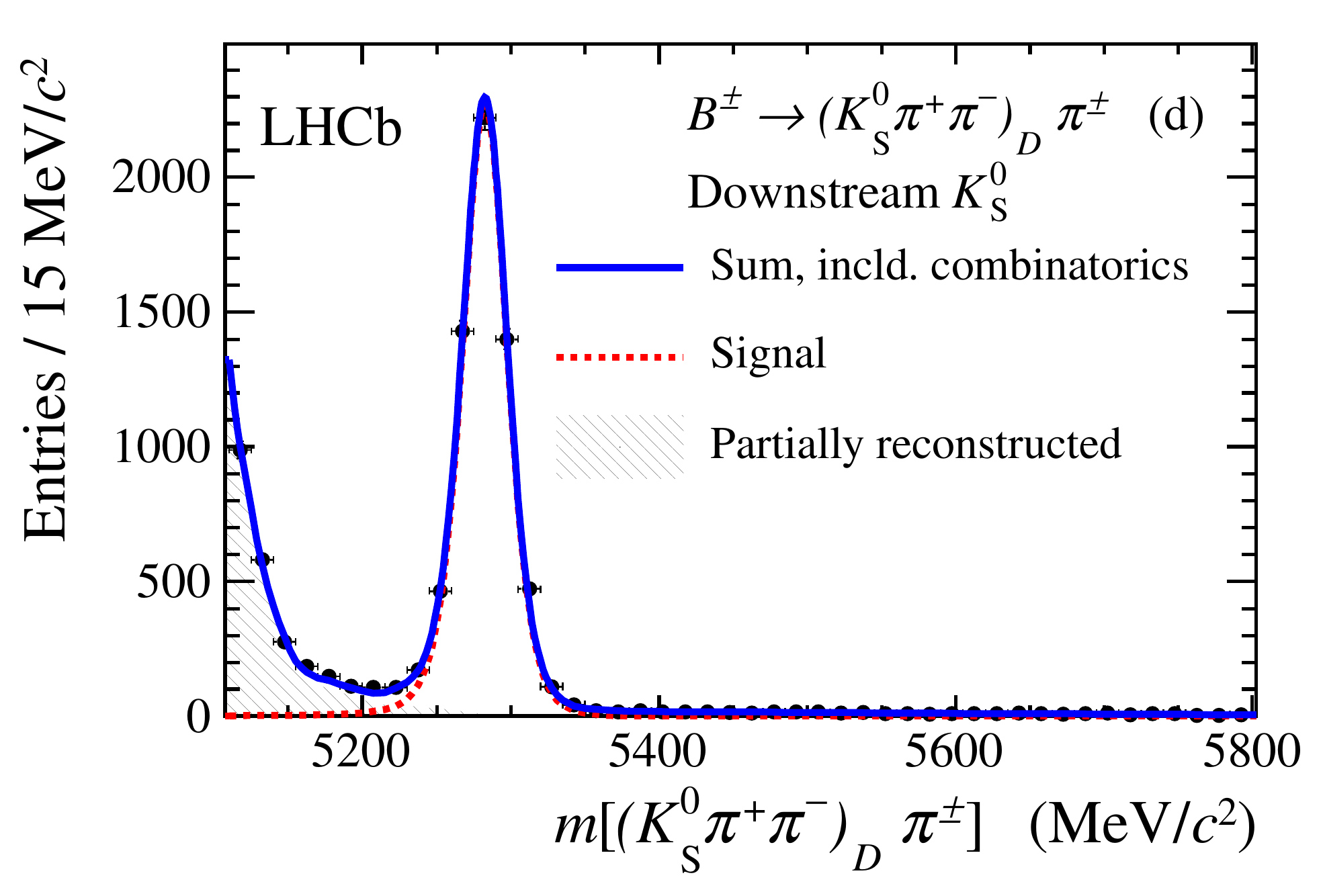}
\caption{\small Invariant mass distributions of (a,c) $B^\pm \to D K^\pm$ and  (b,d) $B^\pm \to D \pi^\pm$ candidates, with $D \to \KS \pi^+\pi^-$, divided between the (a,b) long  and (c,d) downstream \KS categories. Fit results, including the signal and background components, are superimposed.
}
\label{fig:mass_kspipi}
\end{figure}

\begin{figure}[h!]
\centering
\includegraphics[width=0.48\textwidth]{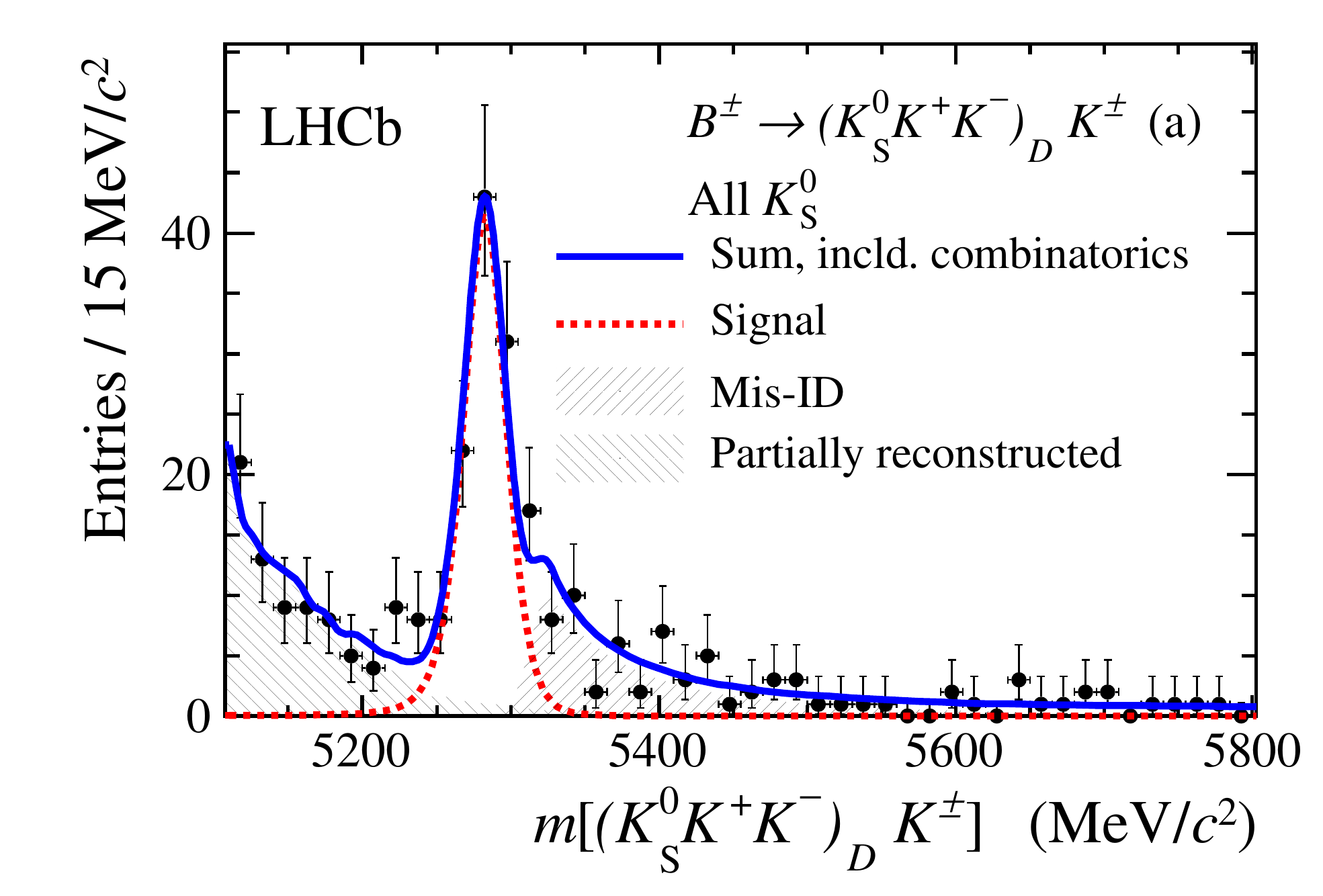}
\includegraphics[width=0.48\textwidth]{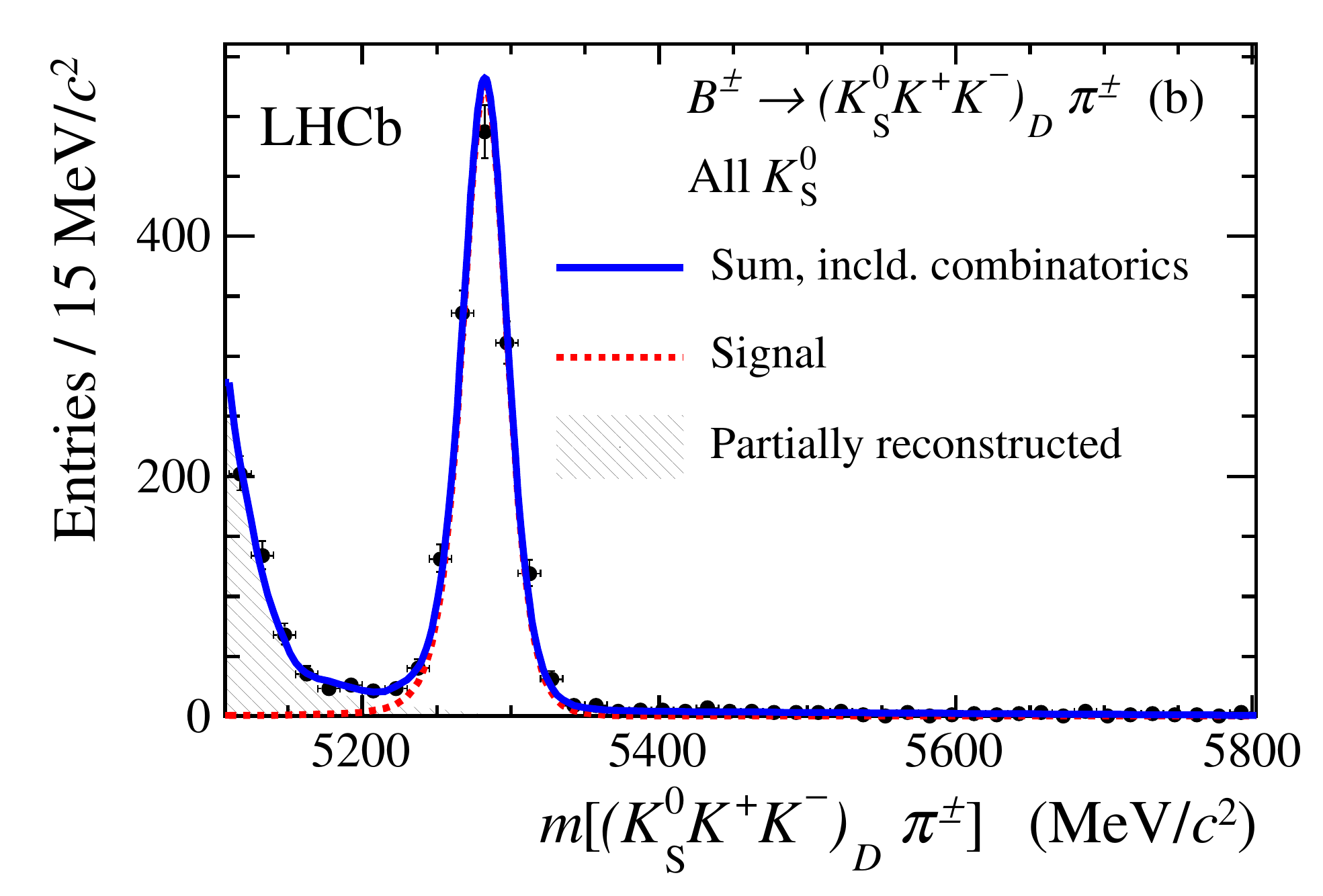}
\caption{\small Invariant mass distributions of (a) $B^\pm \to D K^\pm$ and (b) $B^\pm \to D \pi^\pm$ candidates, with $D \to \KS K^+K^-$, shown with both \KS categories combined.  Fit results, including the signal and background components, are superimposed.  }
\label{fig:mass_kskk}
\end{figure}

The invariant mass distributions of the selected candidates are shown in Fig.~\ref{fig:mass_kspipi} for $B^\pm \to D K^\pm$ and $B^\pm \to D \pi^\pm$, with $D \to \KS \pi^+\pi^-$ decays, divided between the long and downstream \KS categories.    Figure~\ref{fig:mass_kskk} shows the corresponding distributions for final states with $D \to \KS K^+ K^-$, here  integrated over the two \KS categories.
The result of an extended, unbinned, maximum likelihood fit to these distributions is superimposed.   
The fit is performed simultaneously for  $B^\pm \to D K^\pm$ and $B^\pm \to D \pi^\pm$, including both  $D \to \KS \pi^+\pi^-$ and 
$D \to \KS K^+K^-$ decays, allowing several parameters to be different for long and downstream \KS categories.   
The fit range is between 5110~\mevcc and 5800~\mevcc in invariant mass.
At this stage in the analysis the fit does not distinguish between the different regions of Dalitz plot or $B$ meson charge. The purpose of this global fit is to determine the parameters that describe the invariant mass spectrum in preparation for the binned fit described in Sect.~\ref{sec:analysis}.

The signal probability density function (PDF) is a Gaussian function with asymmetric tails where the unnormalised form is given by  
\begin{equation}
f(m; m_0,\alpha_L,\alpha_R,\sigma) = \left\lbrace{\exp[-(m-m_0)^2/(2\sigma^2 + \alpha_L(m-m_0)^2)], m<m_0; \atop \exp[-(m-m_0)^2/(2\sigma^2 + \alpha_R(m-m_0)^2)], m>m_0;}\right.
\end{equation}
where $m$ is the candidate mass, $m_0$ the $B$ mass and $\sigma$, $\alpha_L$, and $\alpha_R$ are free parameters in the fit. The parameter $m_0$ is taken as common for all classes of signal. The parameters describing the asymmetric tails are fitted separately for events with long and downstream  \KS categories. The resolution of the Gaussian function is left as a free parameter for the two \KS categories, but the ratio between this resolution
in  $B^\pm \to D K^\pm$ and $B^\pm \to D \pi^\pm$ decays is required to be the same, independent of category.
The resolution is determined to be around 15~\mevcc  for   $B^\pm \to D \pi^\pm$  decays of both \KS classes,  and is smaller by a factor $0.95 \pm 0.06$ for  $B^\pm \to D K^\pm$. The yield of $B^\pm \to D \pi^\pm$ candidates in each category is determined in the fit. Instead of fitting the yield of the $B^\pm \to D K^\pm$ candidates separately, the ratio $\mathcal{R} = N(B^\pm \to D K^\pm)$/$N(B^\pm \to D \pi^\pm)$ is a free parameter and is common across all categories. 

The background has contributions from random track combinations and partially reconstructed $B$ decays.  The random track combinations are modelled by linear PDFs, the parameters of which are floated separately for each class of decay. 
Partially reconstructed backgrounds are described empirically. Studies of simulated events show that the partially reconstructed backgrounds are dominated by decays that involve a $D$ meson decaying to $\KS h^+h^-$. Therefore the same PDF is used to describe these backgrounds as used in a similar analysis of $\Bpm\to DK^\pm$ decays, with $D \to K^\pm\pi^\mp$, $K^+K^-$ and $\pi^+\pi^-$\cite{LHCBADS}. In that analysis the shape was constructed by applying the selection to a large simulated sample containing many common backgrounds, each weighted by its production rate and branching fraction. The invariant mass distribution for the surviving candidates was corrected to account for small differences in resolution and PID performance between  data and simulation, and two background PDFs were extracted by kernel estimation~\cite{Kernel}; one for $B^\pm \to D K^\pm$ and one for  $B^\pm \to D \pi^\pm$  decays. The partially reconstructed background PDFs are found to give a good description of both \KS categories.

An additional and significant background component exists in the $B^\pm \to D K^\pm$ sample, arising from the dominant  $B^\pm \to D \pi^\pm$ decay on those occasions where the bachelor particle is misidentified as a kaon by the RICH system.  In contrast, the $B^\pm \to D K^\pm$ contamination in the $B^\pm \to D \pi^\pm$ sample can be neglected. The size of this background is calculated through knowledge of PID and misidentification efficiencies, which are obtained from  large samples of kinematically selected $D^{\ast \pm} \to D \pi^\pm$, $D \to K^\mp \pi^\pm$ decays.   The kinematic properties of the particles in the calibration sample are reweighted to match  those of the bachelor particles in the $B$ decay sample, thereby ensuring that the measured PID performance is representative of that in the $B$ decay sample.  The efficiency to identify a kaon correctly  is found to be around 86\%, and that for a pion to be around 96\%.  The misidentification efficiencies are the complements of these numbers.   From this information and from knowledge of the number of  reconstructed $B^\pm \to D \pi^\pm$ decays, the amount of this background surviving the $B^\pm \to D K^\pm$ selection can be determined.
 The invariant mass distribution of the misidentified candidates is described by a Crystal Ball function~\cite{Skwarnicki:1986xj} with the tail on the high mass side, the parameters of which are fitted in common between all the $B^\pm \to D K^\pm$ samples.

The number of $B^\pm \to D K^\pm$ candidates in all categories is determined by $\mathcal{R}$, and the number of $B^\pm \to D \pi^\pm$ events in the corresponding category. The ratio $\mathcal{R}$ is determined in the fit and measured to be 0.085$\pm$0.005 (statistical uncertainty only) and is consistent with that observed in Ref.~\cite{LHCBADS}. The yields returned by the invariant mass fit in the full fit region are scaled to the signal region, defined as 5247--5317~\mevcc, and are presented in Tables~\ref{tab:kspipi_yields} and~\ref{tab:kskk_yields} for the $D \to \KS \pi^+\pi^-$ and  $D \to \KS K^+K^-$ selections respectively.  In the $B^\pm \to D(\KS\pi^+\pi^-)K^\pm$ sample there are $654 \pm 28$ signal candidates, with a purity of 86\%. The corresponding numbers for the  $B^\pm \to D(\KS K^+K^-)K^\pm$ sample are $102 \pm 5$ and 88\%, respectively. The contamination in the $B^\pm \to D K^\pm$  selection receives approximately equal contributions from misidentified $B^\pm \to D \pi^\pm$ decays, combinatoric background and partially reconstructed decays.  The partially reconstructed component in the signal region is dominated by decays of the type $B \to D \rho$, in which a charged pion from the $\rho$ decay is misidentified as the bachelor kaon, and $B^\pm \to D^{*} \pi^\pm$, again with a misidentified pion.

\begin{table}[htb]
\centering
\caption{\small Yields and statistical uncertainties  in the signal region from the invariant mass fit, scaled from the full fit mass range,  for candidates passing the $B^\pm \to D h^\pm$, $D \to \KS \pi^+\pi^-$ selection. Values are shown separately for candidates containing long and downstream \KS decays.
The signal region is between 5247~\mevcc and 5317~\mevcc and the full fit range is between 5110~\mevcc and 5800~\mevcc. \label{tab:kspipi_yields}} \vspace*{0.1cm}
\begin{tabular}{l|cc|cc}
& \multicolumn{2}{c|}{$B^\pm \to D K^\pm$ selection} & \multicolumn{2}{c}{$B^\pm \to D \pi^\pm$ selection} \\ \cline{2-5}
Fit component & Long  &  Downstream  &   Long  &  Downstream \\ \hline

$B^\pm \to D K^\pm$ & $213 \pm 13$ & $441 \pm 25$ & -- & -- \\
$B^\pm \to D \pi^\pm$ & $11 \pm 3$ & $22 \pm 5$ & $2809 \pm 56$ & $5755 \pm 82$ \\
Combinatoric & $\phantom{0}9 \pm 4$ & $29 \pm 6$ & $\phantom{0}22 \pm 3$ & $\phantom{0}90 \pm 7$ \\
Partially reconstructed & $11 \pm 1$ & $25 \pm 2$ & $\phantom{0}25 \pm 1$ & $\phantom{0}55 \pm 1$ \\
\end{tabular}
\end{table}

\begin{table}[htb]
\centering
\caption{\small Yields and statistical uncertainties  in the signal region from the invariant mass fit, scaled from the full fit mass range, for candidates passing the $B^\pm \to D h^\pm$, $D \to \KS K^+K^-$ selection. Values are shown separately for candidates containing long and downstream \KS decays.
The signal region is between 5247~\mevcc and 5317~\mevcc and the full fit range is between 5110~\mevcc and 5800~\mevcc.
\label{tab:kskk_yields}}  \vspace*{0.1cm}
\begin{tabular}{l|cc|cc}
& \multicolumn{2}{c|}{$B^\pm \to D K^\pm$ selection} & \multicolumn{2}{c}{$B^\pm \to D \pi^\pm$ selection} \\ \cline{2-5}
Fit component & Long  &  Downstream  &   Long  &  Downstream \\ \hline

$B^\pm \to D K^\pm$ & $32 \pm 2$ & $70 \pm 4$ & -- & -- \\
$B^\pm \to D \pi^\pm$ & $\phantom{0}1.6 \pm 1.2$ & $\phantom{0}3.4 \pm 1.8$ & $417 \pm 20$ & $913 \pm 29$ \\
Combinatoric & $\phantom{0}0.6 \pm 0.5$ & $\phantom{0}2.5 \pm 0.9$ & $\phantom{0}4.8 \pm 1.4$ & $18 \pm 2$ \\
Partially reconstructed & $\phantom{0}2.2 \pm 0.4$ & $\phantom{0}2.9 \pm 0.5$ & $\phantom{0}3.7 \pm 0.3$ & $\phantom{0}7.7 \pm 0.5$ \\
\end{tabular}
\end{table}

\begin{figure}[htb]
\centering
\includegraphics[width=0.45\textwidth]{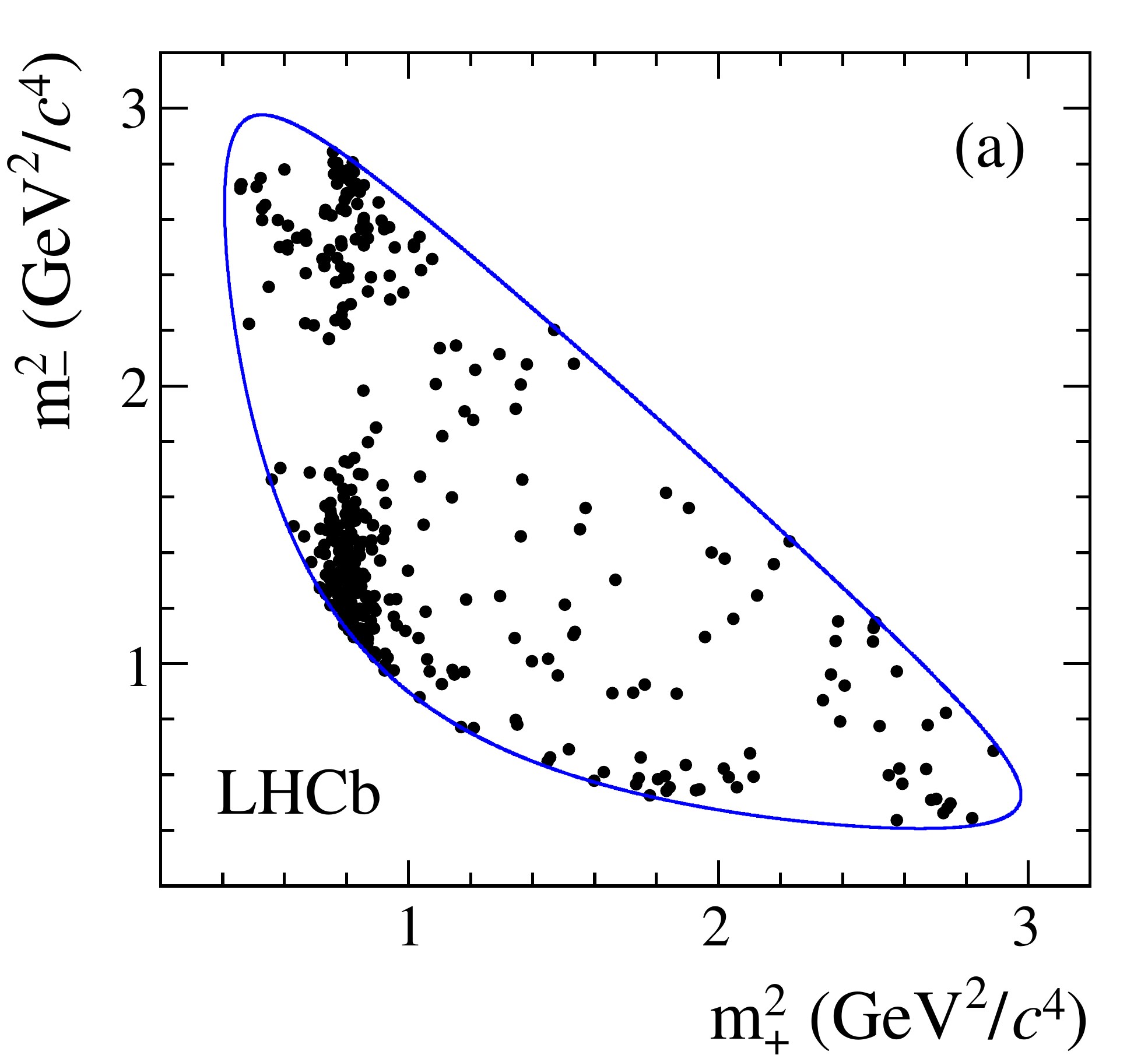}
\includegraphics[width=0.45\textwidth]{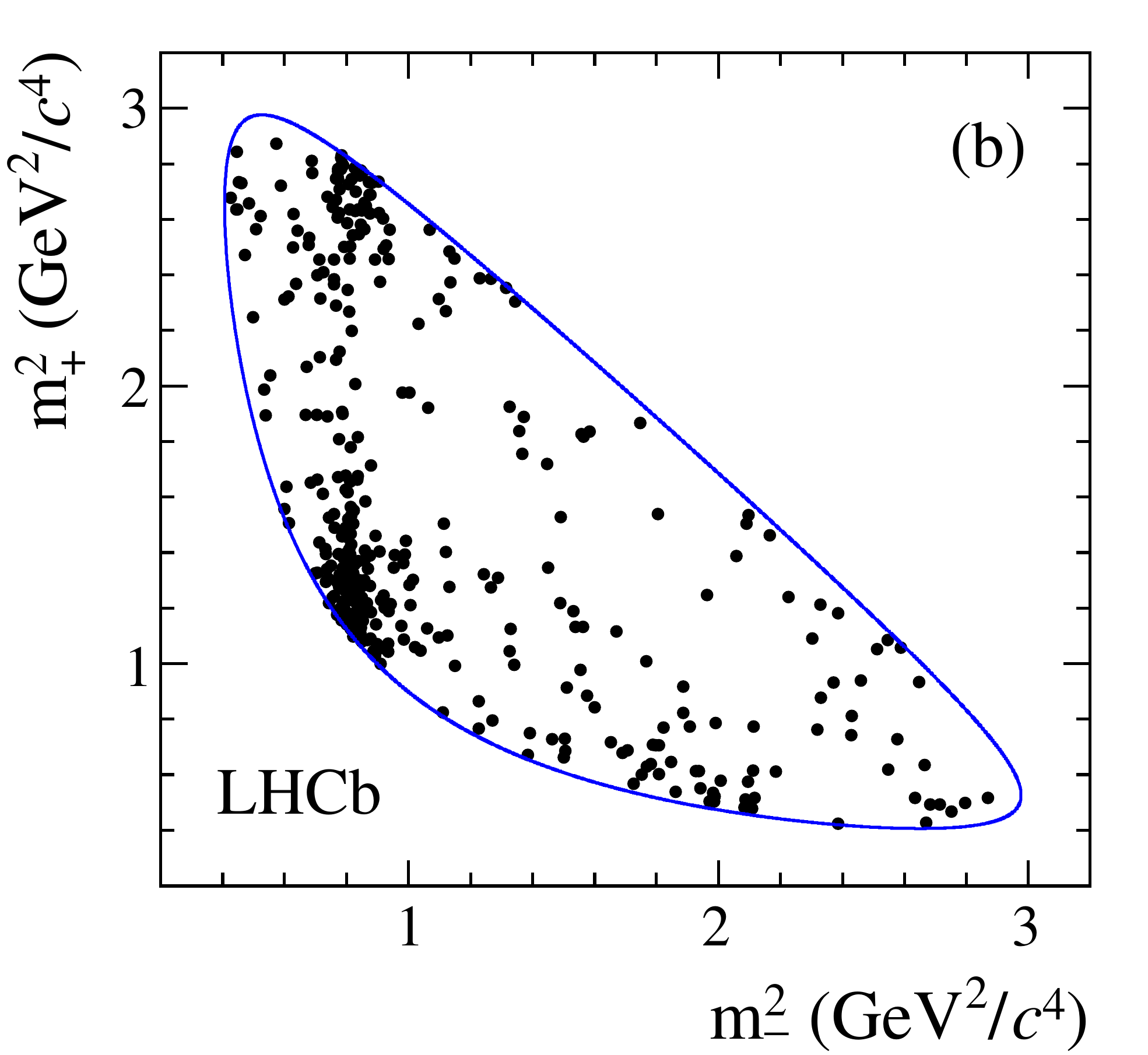}
\includegraphics[width=0.45\textwidth]{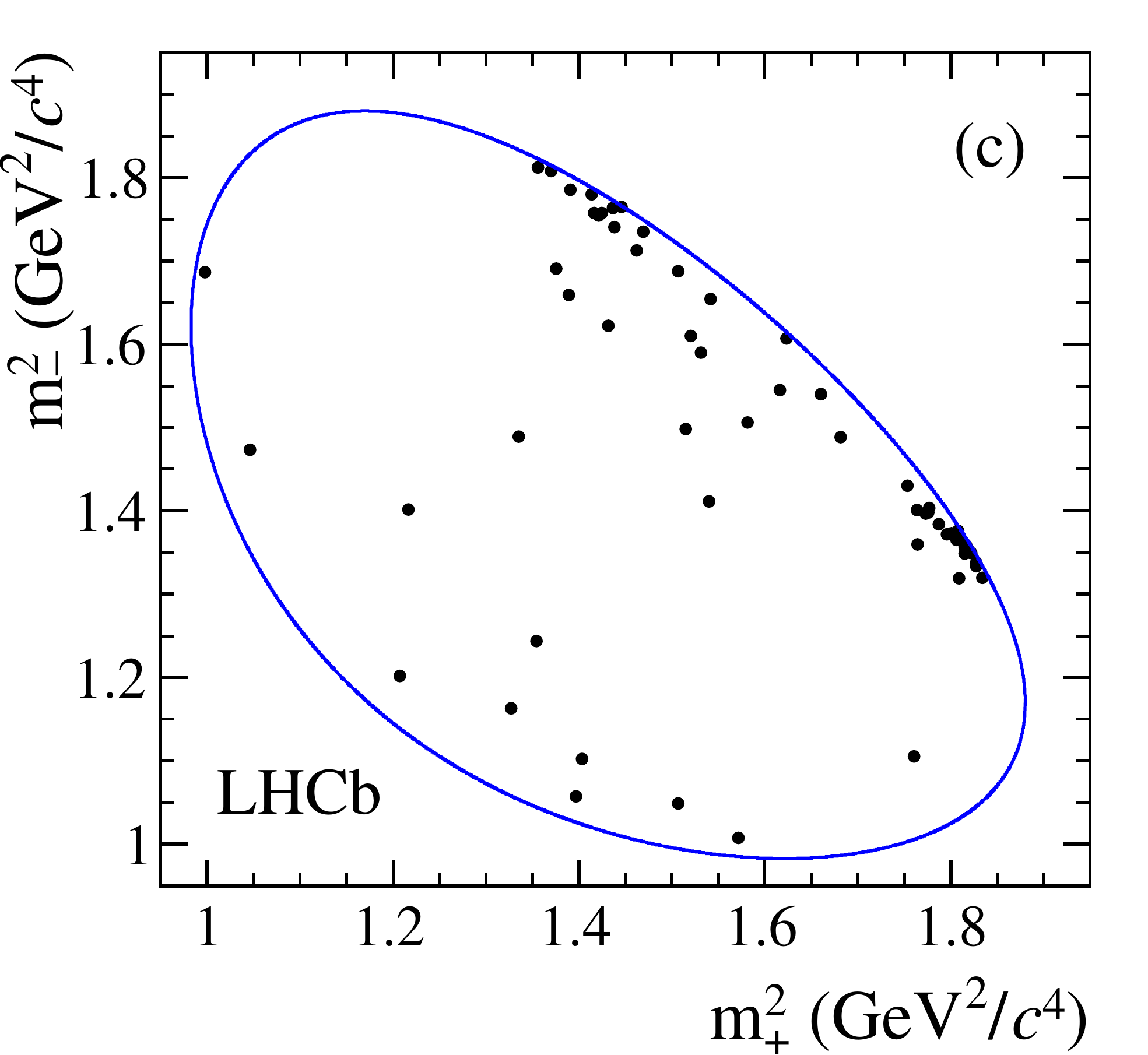}
\includegraphics[width=0.45\textwidth]{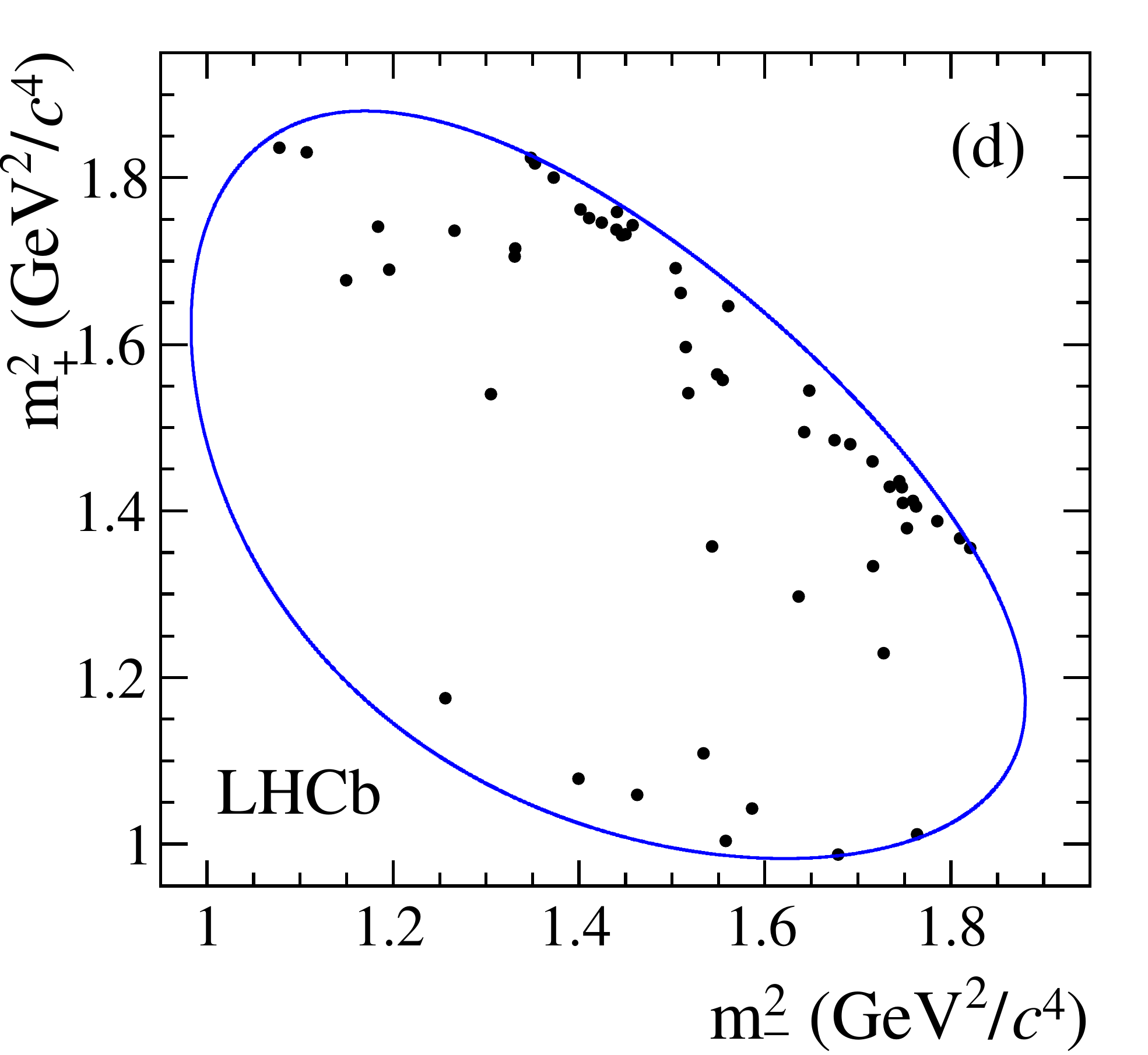}
\caption{\small Dalitz plots of $B^\pm \to D K^\pm$ candidates in the signal region for (a,b) $D \to \KS \pi^+\pi^-$  and  (c,d) $D \to \KS K^+ K^-$ decays,
divided between (a,c) $B^+$ and (b,d) $B^-$. The boundaries of the kinematically-allowed regions are also shown.}
\label{fig:dalitz}
\end{figure}

The Dalitz plots for $B^\pm \to DK^\pm$ data in the signal region for the two $D \to \KS h^+h^-$ final states are shown in Fig.~\ref{fig:dalitz}.  Separate plots are shown for $B^+$ and $B^-$ decays.


\section{Binned Dalitz fit}
\label{sec:analysis}

The purpose of the binned Dalitz plot fit is to measure the  \CP-violating parameters $x_{\pm}$ and $y_{\pm}$, as introduced in Sect.~\ref{sec:formalism}.  Following Eq.~(\ref{eq:populations}) these parameters can be determined from the populations of each 
$B^\pm \to D K^\pm$ Dalitz plot bin given the external information that is available for the $c_i$, $s_i$ and $K_i$ parameters. 
 In order to know the signal population in each bin it is necessary both to subtract background and to correct for acceptance losses from the trigger, reconstruction and selection.  

Although  the absolute numbers of $B^+$ and $B^-$  decays integrated over the Dalitz plot have some dependence on $x_{\pm}$ and $y_{\pm}$, the additional sensitivity gained compared to using just the relative bin-to-bin yields is negligible, and is therefore not used. Consequently the analysis is insensitive to any $B$ production asymmetries, and only knowledge of the  relative acceptance  is required.  The relative acceptance is determined from the control channel $B^\pm \to D \pi^\pm$.  In this decay the ratio of $b \to u\bar{c}d$ to $b \to c\bar{u}d$ amplitudes is expected to be very small ($\sim 0.005$) and thus, to a good approximation, interference between the transitions can be neglected.  Hence the relative population of decays expected in each $B^\pm \to D \pi^\pm$ Dalitz plot bin can be predicted using the $K_i$ values calculated with the $D \to \KS h^+h^-$ model.  Dividing the background-subtracted yield observed in each bin by this prediction enables the relative acceptance to be determined, and then applied to the  
$B^\pm \to D K^\pm$ data. 
In order to optimise the statistical precision of this procedure, the bins $+i$ and $-i$ are combined in the calculation,
since the efficiencies in these symmetric regions are expected to be the same in the limit that there are no 
charge-dependent reconstruction asymmetries.  
It is found that the variation in relative acceptance between non-symmetric bins is at most $\sim 50\%$, with the 
lowest efficiency occurring in those regions where one of the pions has low momentum.

Separate fits are performed to the $B^+$ and $B^-$ data.  Each fit simultaneously considers the two \KS categories, the $\Bpm \to D \Kpm$ and $\Bpm \to D \pipm$ candidates, and the two $D \to \KS h^+h^-$  final states.   In order to assess the impact of the  $D \to \KS K^+K^-$ data the fit is then repeated including only the   $D \to \KS \pi^+\pi^-$ sample.
The PDF parameters for both the signal and background invariant mass distributions are fixed to the values determined in the global fit.   The yields of all the background contributions in each bin are free parameters, apart from bins where a very low contribution is determined from an initial fit, in which case they are fixed to zero, to facilitate the calculation of the error matrix.  The yields of signal candidates for each bin in the $\Bpm\to D \pipm$ sample are also free parameters.   The amount of signal in each bin for the $\Bpm\to D \Kpm$ sample is determined by varying the integrated yield and the  $x_{\pm}$ and $y_{\pm}$ parameters.

A large ensemble of simulated experiments are performed to validate the fit procedure.  In each experiment the number and distribution of signal and background candidates are generated according to the expected distribution  in data, and the full fit procedure is then executed.  The values for $x_{\pm}$ and $y_{\pm}$ are set close to those determined by previous measurements~\cite{HFAG}.   It is found from this exercise that the errors are well estimated.   Small biases are, however, observed in the central values returned by the fit and these are applied as corrections to the results obtained on data. 
The bias is $(0.2-0.3)\times 10^{-2}$ for most parameters but rises to $1.0 \times 10^{-2}$ for $y_+$. This bias is due to the low yields in some of the bins and is an inherent feature of the maximum likelihood fit. This behaviour is associated with the size of data set being fit, since when simulated experiments are performed with larger sample sizes the biases are observed to reduce.

The results of the fits are presented in Table~\ref{tab:theresults}.  The systematic uncertainties are discussed in Sect.~\ref{sec:syst}.    The statistical uncertainties are compatible with those predicted by simulated experiments.
 The inclusion of the $D \to \KS K^+K^-$ data improves the precision on $x_{\pm}$ by around 10\%, and has little impact on $y_{\pm}$.
This behaviour is expected, as the measured values of $c_i$ in this mode, which multiply $x_\pm$ in Eq.~(\ref{eq:xydefinitions}), are significantly larger than those of $s_i$, which multiply $y_\pm$.
The two sets of results are compatible within the statistical and uncorrelated systematic uncertainties.  

\begin{table}[tb]
\centering
\caption{Results for $x_{\pm}$ and $y_{\pm}$ from the fits to the  data in the case when
both  $D \to \KS \pi^+\pi^-$ and $D \to \KS K^+K^-$ are considered  and when only 
the $D \to \KS \pi^+\pi^-$ final state is included.  The first, second, and third uncertainties are the statistical, the experimental systematic, and the error associated with the precision of the strong-phase parameters, respectively.
The correlation coefficients are calculated including all sources of uncertainty  (the values in parentheses correspond to the case where only the statistical uncertainties are considered).}
\label{tab:theresults} \vspace*{0.1cm}
\begin{tabular}{r|rr} 
Parameter &  \multicolumn{1}{c}{All data} &  \multicolumn{1}{c}{$D \to \KS  \pi^+\pi^- $ alone} \\ \hline
$x_-$ [$\times 10^{-2}$] &   $0.0 \pm 4.3 \pm 1.5 \pm 0.6$   &  $1.6 \pm 4.8 \pm 1.4 \pm 0.8$ \\
$y_-$ [$\times 10^{-2}$] &   $2.7 \pm 5.2 \pm 0.8 \pm 2.3$   &  $1.4 \pm 5.4 \pm 0.8 \pm 2.4$ \\
corr($x_-$,$y_-$)  &  $-0.10$ ($-0.11$) & $-0.12$ ($-0.12$) \\
$x_+$ [$\times 10^{-2}$] &   $-10.3 \pm 4.5 \pm 1.8 \pm 1.4$  &  $-8.6 \pm 5.4 \pm 1.7 \pm 1.6$ \\
$y_+$ [$\times 10^{-2}$] &   $-0.9 \pm 3.7 \pm 0.8 \pm 3.0$  &  $-0.3 \pm 3.7 \pm 0.9 \pm 2.7$ \\
corr($x_+$,$y_+$)  &  0.22 (0.17) & $0.20$ (0.17) \\
\end{tabular}
\end{table}

\begin{figure}[t]
\begin{center}
\includegraphics[width=0.48\textwidth]{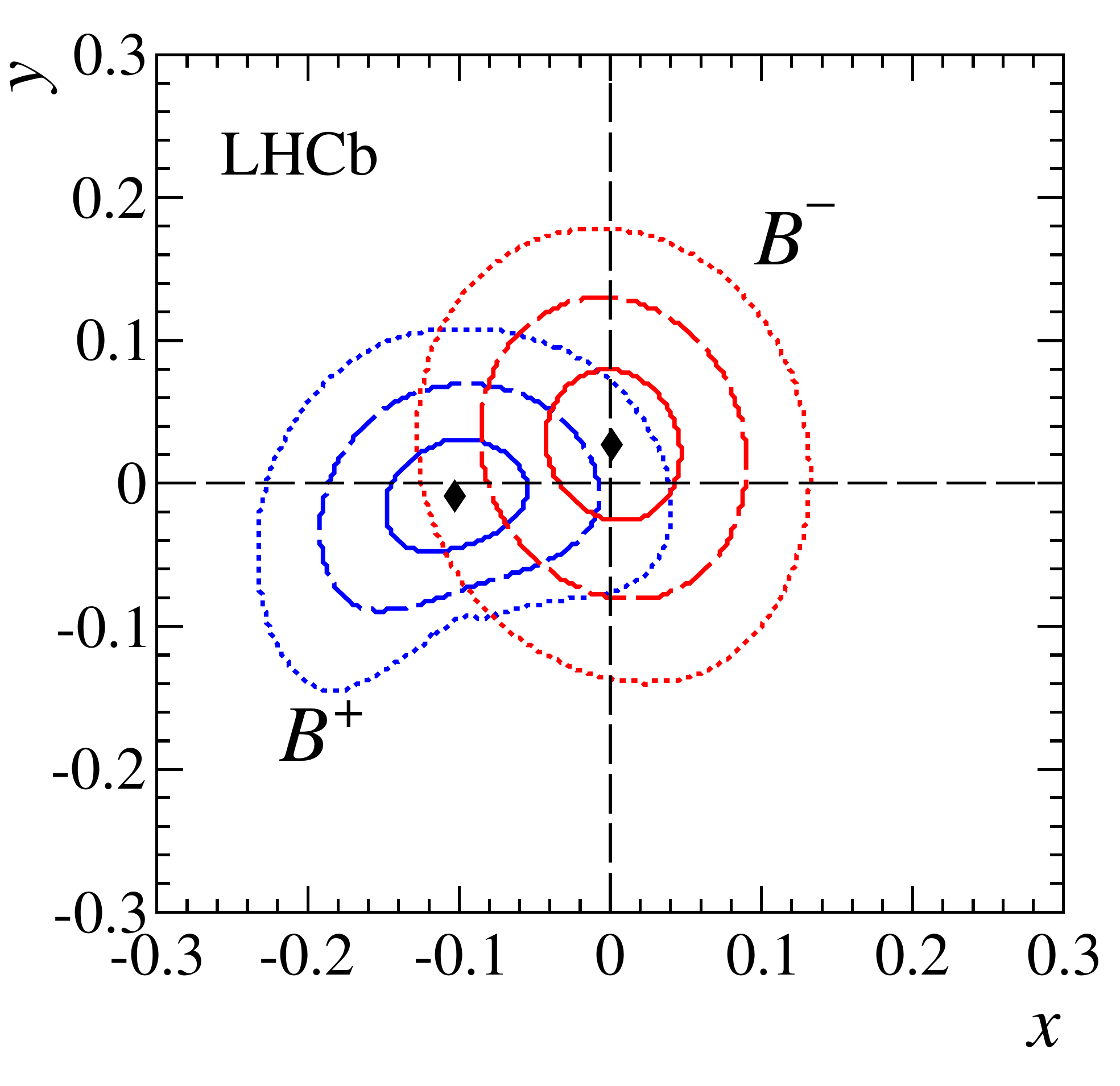}
\caption{\small One (solid), two (dashed) and three (dotted) standard deviation confidence levels for  $(x_+,y_+)$ (blue) and $(x_{-}, y_{-})$ (red) as measured in $B^\pm \to D K^\pm$ decays (statistical only). The points represent the best fit central values.}
\label{fig:sunnysideup}
\end{center}
\end{figure}

The measured values of $(x_{\pm}, y_{\pm})$ from the fit to all data, with their statistical likelihood contours are shown in Fig.~\ref{fig:sunnysideup}.   The expected signature for a sample that exhibits \CP-violation is that the two vectors defined by the coordinates $(x_-,y_-)$ and $(x_+,y_+)$ should both be non-zero in magnitude, and have different phases.  The data show this behaviour, but are also compatible with the no \CP violation hypothesis.

\begin{figure}[htb]
\centering
\includegraphics[width=0.45\textwidth]{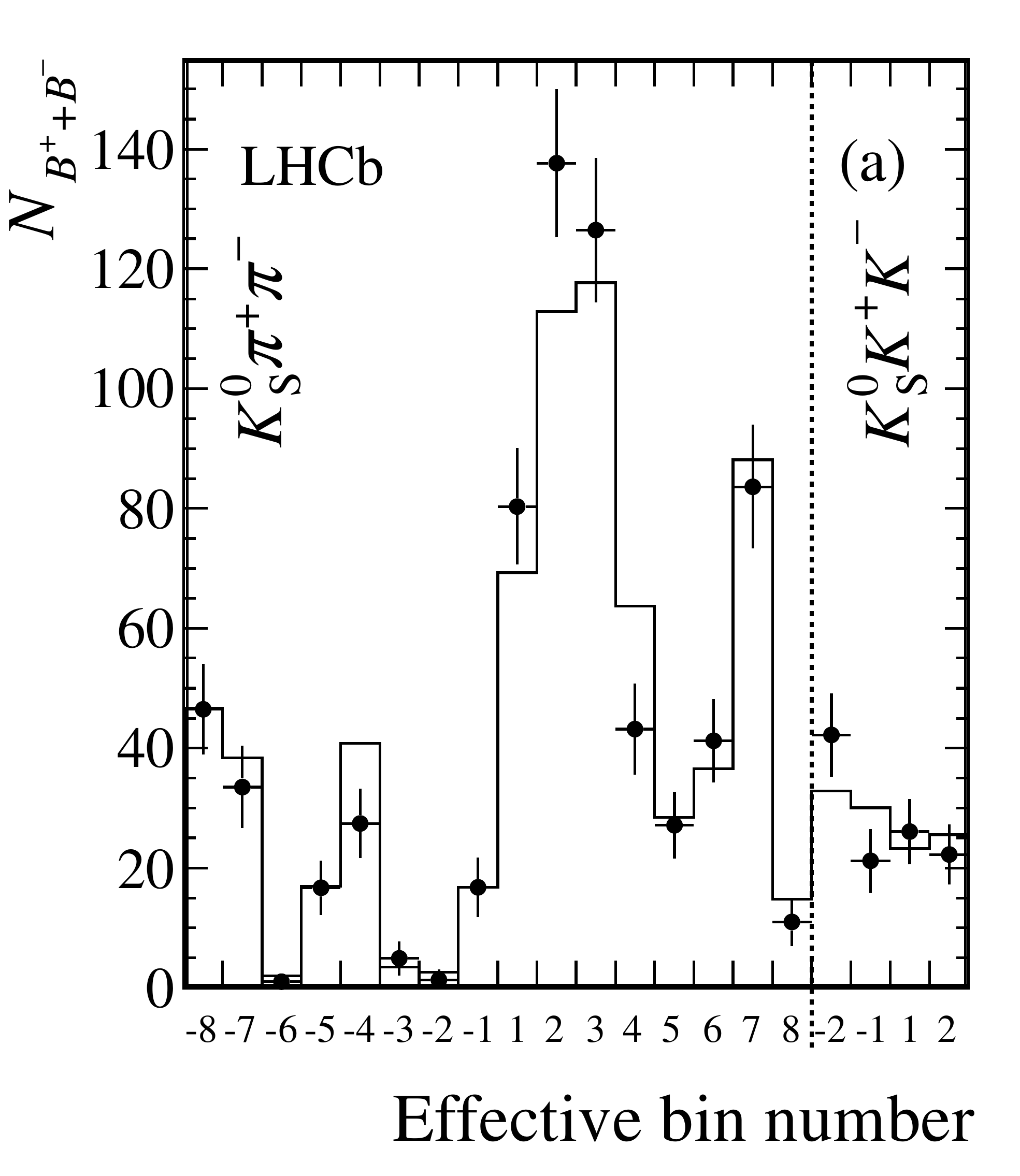}
\includegraphics[width=0.45\textwidth]{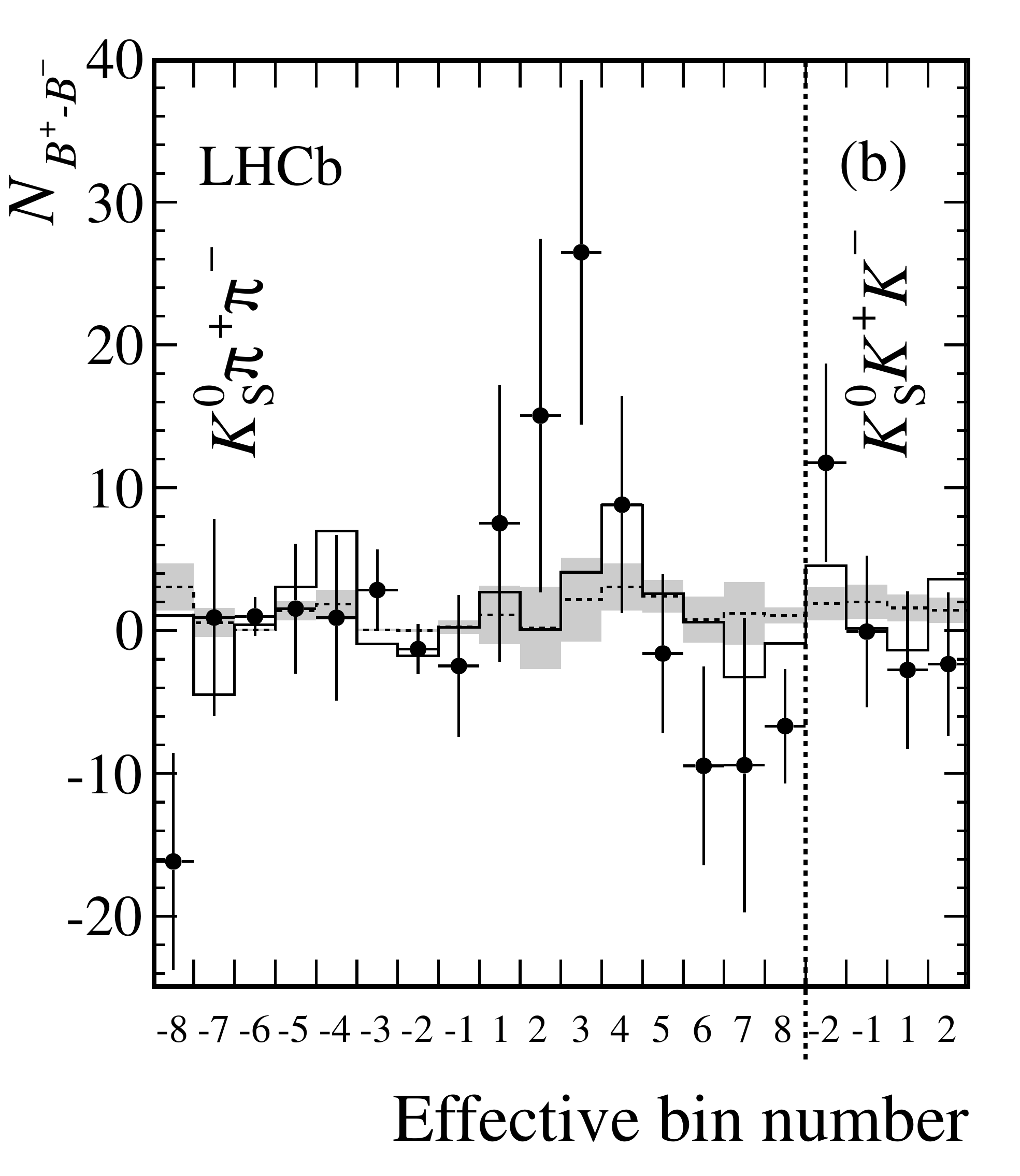}
\caption{\small 
Signal yield in effective bins compared with prediction of $(x_\pm, y_\pm)$ fit (black histogram) for $D \to \KS \pi^+\pi^-$ 
and $D \to \KS K^+K^-$. Figure (a) shows the sum of $B^+$ and $B^-$ yields. Figure (b) shows the difference of $B^+$ and $B^-$ yields. Also shown (dashed line and grey shading) is the expectation and uncertainty for the zero \CP-violation hypothesis. }
\label{fig:bin_plots}
\end{figure}

In order to investigate whether the binned fit gives an adequate description of the data,
a study is performed to compare the observed number of signal candidates in each bin  with that expected given the fitted total yield and values of $x_\pm$ and $y_\pm$.  The number of signal candidates is determined by fitting in each bin for the $B^\pm \to D K^\pm$ contribution 
for long and downstream \KS decays combined,  with no assumption on how this component is distributed over the Dalitz plot.  
Figure~\ref{fig:bin_plots} shows the results in effective bin number separately for $N_{B^+ + B^-}$, the sum of $B^+$ and $B^-$ candidates, which is a \CP-conserving observable, and for the difference $N_{B^+ - B^-}$, which is sensitive to \CP violation.  The effective bin number is equal to the normal bin number for $B^+$, but is defined to be this number multiplied by $-1$ for $B^-$.
The expectations from the ($x_\pm$, $y_\pm$) fit  are superimposed  as is, for the $N_{B^+ - B^-}$ distribution, the prediction for the case
$x_\pm = y_\pm = 0$.  Note that the zero \CP violation prediction is not a horizontal  line at $N_{B^+ -  B^-} =0$ because it is calculated using the total $B^+$ and $B^-$ yields from the full fit, and using bin efficiencies that are determined separately for each sample.
The data and fit expectations are compatible for both distributions yielding 
a $\chi^2$ probability of 10\%  for $N_{B^+ + B^-}$ and 
34\%  for $N_{B^+ - B^-}$. 
The results for the $N_{B^+ - B^-}$ distribution are also compatible with the no \CP-violation hypothesis  ($\chi^2$ probability = 16\% ).

\section{Systematic uncertainties}
\label{sec:syst}

Systematic uncertainties are evaluated for the fits to the full data sample and are presented in Table~\ref{tab:syst_all}. In order to
understand the impact of the CLEO-c $(c_i, s_i)$ measurements the errors arising from this source are kept separate from the other experimental uncertainties.  Table~\ref{tab:syst_kspipi} shows the uncertainties for the case  where only $D \to \KS \pi^+\pi^-$  decays are included.    Each contribution to the systematic uncertainties is now discussed in turn.

\begin{table}[htb]
\centering
\caption{\small Summary of statistical, experimental and strong-phase uncertainties on $x_\pm$ and $y_\pm$ in the case where
both  $D \to \KS \pi^+\pi^-$ and $D \to \KS K^+K^-$ decays are included in the fit. All entries are given in multiples of $10^{-2}$.}\label{tab:syst_all} \vspace*{0.1cm}
\begin{tabular}{l|rrrr}
Component & $\sigma(x_-)$ & $\sigma(y_-)$ & $\sigma(x_+)$ & $\sigma(y_+)$ \\ \hline
Statistical &                                              $4.3$ & $5.2$ & $4.5$ & $3.7$ \\
& & & & \\
Global fit shape parameters &                                      $0.4$ &  $0.4$ &  $0.6$ &  $0.4$ \\
Efficiency effects &                                                      $0.3$ &  $0.4$ &  $0.3$ &  $0.4$ \\
\CP violation in control mode &                                            $1.3$ &  $0.4$ &  $1.5$ &  $0.2$ \\
Migration &                                                                   $0.4$ &  $0.2$ &  $0.4$ &  $0.2$ \\
Partially reconstructed background &                           $0.2$ &  $0.3$ &  $0.2$ &  $0.2$ \\
PID efficiency &                                                     $0.1$ &  $0.2$ &  $0.2$ &  $<0.1$ \\
Shape of misidentified $B^\pm \to D \pi^\pm$  &  
                                                                                    $0.1$ &  $0.1$ &  $0.3$ &  $<0.1$ \\
Bias correction &                                                          $0.2$ &  $0.3$ &  $0.2$ &  $0.5$ \\
\hline
Total experimental systematic&                                                     $1.5$ & $0.9$ & $1.8$ & $0.8$  \\
& & & & \\
Strong-phase systematic &                                              $0.6$ &  $2.3$ &  $1.4$ &  $3.0$ \\
\end{tabular}
\end{table}

\begin{table}[htb]
\centering
\caption{\small Summary of statistical, experimental and strong-phase uncertainties on $x_\pm$ and $y_\pm$ in the case where
only  $D \to \KS \pi^+\pi^-$ decays are included in the fit. All entries are given in multiples of $10^{-2}$.}\label{tab:syst_kspipi}
\vspace*{0.1cm}
\begin{tabular}{l|rrrr}
Component & $\sigma(x_-)$ & $\sigma(y_-)$ & $\sigma(x_+)$ & $\sigma(y_+)$ \\ \hline
Statistical &                                                                     $4.8$ & $5.4$ & $5.4$ & $3.7$ \\
& & & & \\
Global fit shape parameters &                                      $0.4$ &  $0.4$ &  $0.6$ &  $0.4$ \\
Efficiency effects &                                                       $0.2$ &  $0.2$ &  $0.3$ &  $0.4$ \\
\CP violation in control mode &                                           $1.2$ &  $0.5$ &  $1.5$ &  $0.2$ \\
Migration &                                                                   $0.4$ &  $0.2$ &  $0.4$ &  $0.2$ \\
Partially reconstructed background &                          $0.1$ &  $0.1$ &  $0.3$ &  $0.2$ \\
PID efficiency &                                                         $<0.1$ &  $0.2$ &  $<0.1$ &  $<0.1$ \\
Shape of misidentified $B^\pm \to D \pi^\pm$ &  
                                                                                 $0.1$ &  $<0.1$ &  $0.1$ &  $<0.1$ \\
Bias correction &                                                         $0.2$ &  $0.3$ &  $0.2$ &  $0.6$ \\
\hline
Total experimental systematic&                                                    $1.4$ & $0.8$ & $1.7$ & $0.9$  \\
& & & & \\
Strong-phase systematic &                                              $0.8$ &  $2.4$ &  $1.6$ &  $2.7$ \\
\end{tabular}
\end{table}


The uncertainties on the shape parameters of the invariant mass distributions as determined from the global fit when propagated through to the binned analysis induce uncertainties on $x_\pm$ and $y_\pm$.   In addition, consideration is given to certain assumptions made in the fit.
 For example, the slope of the combinatoric background in the data set containing $D \to \KS K^+ K^-$ decays is fixed to be zero on account of the limited sample size.   The induced errors associated with these assumptions are evaluated and found to be small compared to those coming from the parameter uncertainties themselves, which vary between $0.4 \times 10^{-2}$ and $0.6 \times 10^{-2}$ for the fit to the full data sample. 


The analysis assumes an efficiency that is flat across each Dalitz plot bin.   In reality the efficiency varies, and this leads to a potential bias in the determination of $x_\pm$ and $y_\pm$, since the non-uniform acceptance means that the values of $(c_i, s_i)$ appropriate
for the analysis can differ from those corresponding to the flat-efficiency case.
The possible size of this effect is evaluated in LHCb simulation by dividing each Dalitz plot bin into many smaller cells, and using the \babar amplitude model~\cite{BABAR2008,BABAR2010} to calculate the values of $c_i$ and $s_i$ within each cell.   These values are then averaged together, weighted by the population of each cell after efficiency losses, to obtain an effective $(c_i, s_i)$ for the bin as a whole, and the results compared with those determined assuming a flat efficiency.  The differences between the two sets of results are found to be small compared with the CLEO-c measurement uncertainties.  The data fit is then rerun many times, and the input values of $(c_i, s_i)$ are smeared according to the size of these differences, and the mean shifts are assigned as a systematic uncertainty.  These shifts vary between $0.2 \times 10^{-2}$ and $0.3 \times 10^{-2}$.


The relative efficiency in each Dalitz plot bin is determined from the  $B^\pm \to D \pi^\pm$ control sample.  Biases can enter the 
measurement if there are differences in the relative acceptance over the Dalitz plot between the control sample and that of signal $B^\pm \to D K^\pm$ decays.  Simulation studies show that the acceptance shapes are very similar between the two decays,  but small variations exist which can be attributed to kinematic correlations induced by the different PID requirements on the bachelor particle from the $B$ decay.  When included in the data fit, these variations induce biases that vary between $0.1 \times 10^{-2}$ and $0.3\times 10^{-2}$.  In addition, a check is performed in which the control sample is fitted without combining together bins $+i$ and $-i$ in the efficiency calculation. As a result of this study small uncertainties of $\le 0.3\times 10^{-2}$ are assigned for the $D \to \KS K^+K^-$ measurement  to account for possible biases induced by the difference in interaction cross-section for $K^-$ and $K^+$ mesons interacting with the detector material.
These contributions are combined together with the uncertainty arising from efficiency variation within a Dalitz plot bin to give the component labelled `Efficiency effects' in Tables~\ref{tab:syst_all} and~\ref{tab:syst_kspipi}.


The use of the control channel to determine the relative efficiency on the Dalitz plot assumes that the amplitude of the suppressed tree diagram is negligible. If this is not the case then the $B^-$ final state will receive a contribution from \Dzb decays, and this will lead to the presence of \CP violation via the same mechanism as in $B \to DK$ decays.
The size of any \CP violation that exists in this channel is governed by $r^{D\pi}_B$, $\gamma$ and $\delta^{D\pi}_B$, where the parameters with superscripts are analogous to their counterparts in  $B^\pm \to DK^\pm$ decays.  The naive expectation  is that $r^{D\pi}_B \sim 0.005$ but larger values are possible, and the studies reported in Ref.~\cite{LHCBADS} are compatible with this possibility. 
Therefore  simulated experiments are performed with finite \CP violation injected  in the control channel, conservatively setting  $r^{D\pi}_B$ to be 0.02,
taking a wide variation  in the value of the unknown strong-phase difference $\delta^{D\pi}_B$, and choosing  $\gamma = 70^\circ$.
The experiments are fit under the no \CP violation hypothesis and the largest shifts observed are assigned as a systematic uncertainty.  This contribution is the largest source of experimental systematic uncertainty in the measurement, for example contributing an error of $1.5 \times 10^{-2}$ in the case of $x_+$ in the full data fit. 


The resolution of each decay on the Dalitz plot is approximately 0.004~\gevgevcccc for candidates with long \KS decays and 0.006~\gevgevcccc for those containing downstream \KS in the $m^2_+$ and $m_-^2$ directions.  This is small compared to the typical width of a bin, nonetheless some net migration is possible away from the more densely populated bins.  At first order this effect is accounted for by use of the control channel, but residual effects enter because of the different distribution in the Dalitz plot of the signal events.  Once more a series of simulated experiments is performed to assess the size of any possible bias which is found to vary between $0.2 \times 10^{-2}$ and $0.4 \times 10^{-2}$.

The distribution of the partially reconstructed background is varied over the Dalitz plot according to the uncertainty in the make-up of this background component. From these studies an uncertainty of $(0.2-0.3) \times 10^{-2}$ is assigned to the fit parameters in the full data fit.


Two systematic uncertainties are evaluated that are  associated with the misidentified $B^\pm \to D \pi^\pm$ background in the $B^\pm \to D K^\pm$ sample.  Firstly, there is a $0.2 \times 10^{-2}$ uncertainty on the knowledge of the efficiency of the PID cut that distinguishes pions from kaons.
This is found to have only a small effect
 on the measured values of $x_\pm$ and $y_\pm$.
Secondly, it is possible that the invariant mass distribution of the misidentified background  is not constant over the Dalitz plot, as is assumed in the fit.  This can occur through kinematic correlations between  the reconstruction efficiency on the Dalitz plot of the $D$ decay 
and the momentum of the bachelor pion from the \Bpm decay.  Simulated experiments are performed with different shapes input according to the Dalitz plot bin and the results of simulation studies, and these experiments are then fitted assuming a uniform shape, as in data.  Uncertainties are assigned in the range $(0.1 - 0.3) \times 10^{-2}$.

An uncertainty is assigned to each parameter to accompany the correction that is applied  for the small  bias which is present in the fit procedure.  These uncertainties are determined by performing sets of simulated experiments, in each of which different values of $x_\pm$ and $y_\pm$ are input, corresponding to a range that is wide compared to the current experimental knowledge, and also encompassing the results of this analysis.  The spread in observed bias is taken as the systematic error, and is largest for $y_+$, reaching a value of $0.5 \times 10^{-2}$ in the full data fit.

Finally, several robustness checks are conducted to assess the stability of the results.   These include repeating the analysis with alternative binning schemes for the $D \to \KS \pi^+\pi^-$ data and performing the fits without making any distinction between \KS category.  These tests return results compatible with the baseline procedure.

The total experimental systematic uncertainty from LHCb-related sources is determined to be $1.5 \times 10^{-2}$ on $x_-$, $0.9 \times 10^{-2}$ on $y_-$, $1.8 \times 10^{-2}$ on $x_+$ and $0.8 \times 10^{-2}$ on $y_+$.   These are all smaller than the corresponding statistical uncertainties.  The dominant contribution arises from allowing for the possibility of \CP violation in the control channel, $B \to D \pi$.  In the future, when larger data sets are analysed, alternative analysis methods will be explored to eliminate this potential source of bias.


The limited precision on $(c_i, s_i)$ coming from the CLEO-c measurement induces uncertainties on $x_\pm$ and $y_\pm$~\cite{CLEOCISI}.
These uncertainties are evaluated by rerunning the data fit many times, and smearing the input values of $(c_i, s_i)$ according to their 
measurement errors and correlations.  Values of $(0.6 - 3.0) \times 10^{-2}$ are found for the fit to the full sample.
 When evaluated for the $D \to \KS \pi^+\pi^-$ data set alone, the results are similar in magnitude, but not identical, to those reported in the corresponding Belle analysis~\cite{BELLEMODIND}.  Differences are to be expected, as these uncertainties have a dependence on the central values of the $x_\pm$ and $y_\pm$ parameters, and are sample-dependent for small data sets.   Simulation studies indicate that these uncertainties will be reduced when larger $\Bpm\to D\Kpm$ data sets are analysed. 


After taking account of all sources of uncertainty the correlation coefficient between $x_-$ and $y_-$ in the full fit is calculated to be $-0.10$ and that between $x_+$ and $y_+$ to be $0.22$.  The correlations between $B^-$ and $B^+$ parameters are found to be small and can be neglected.  These correlations are summarised in Table~\ref{tab:theresults}, together with those coming from the statistical uncertainties alone, and those from the fit to $D \to \KS\pi^+\pi^-$ data.


\section{Interpretation}
\label{sec:discussion}

\begin{figure}[htbp]
\begin{center} 
\includegraphics[width=0.45\textwidth]{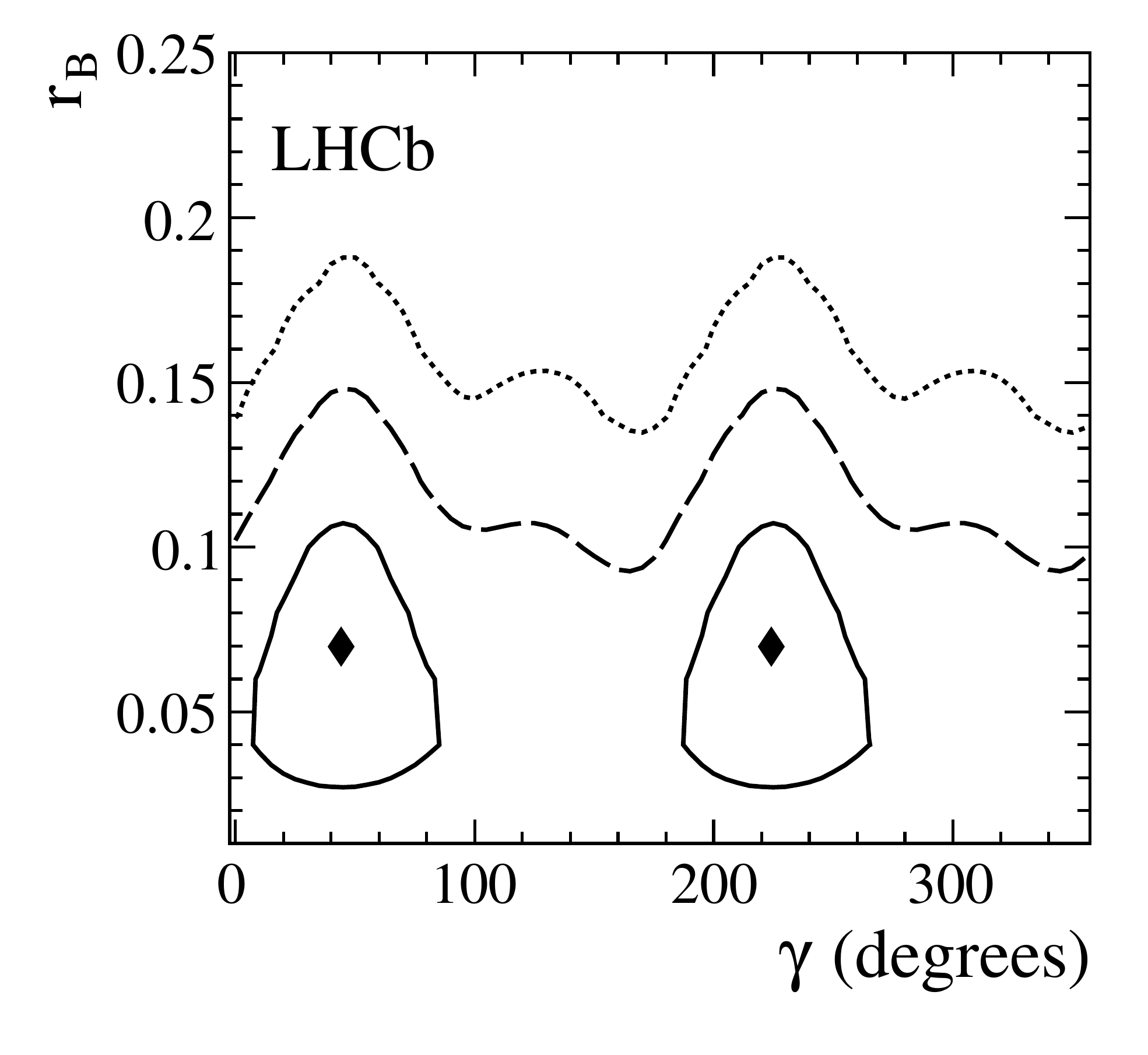}
\includegraphics[width=0.45\textwidth]{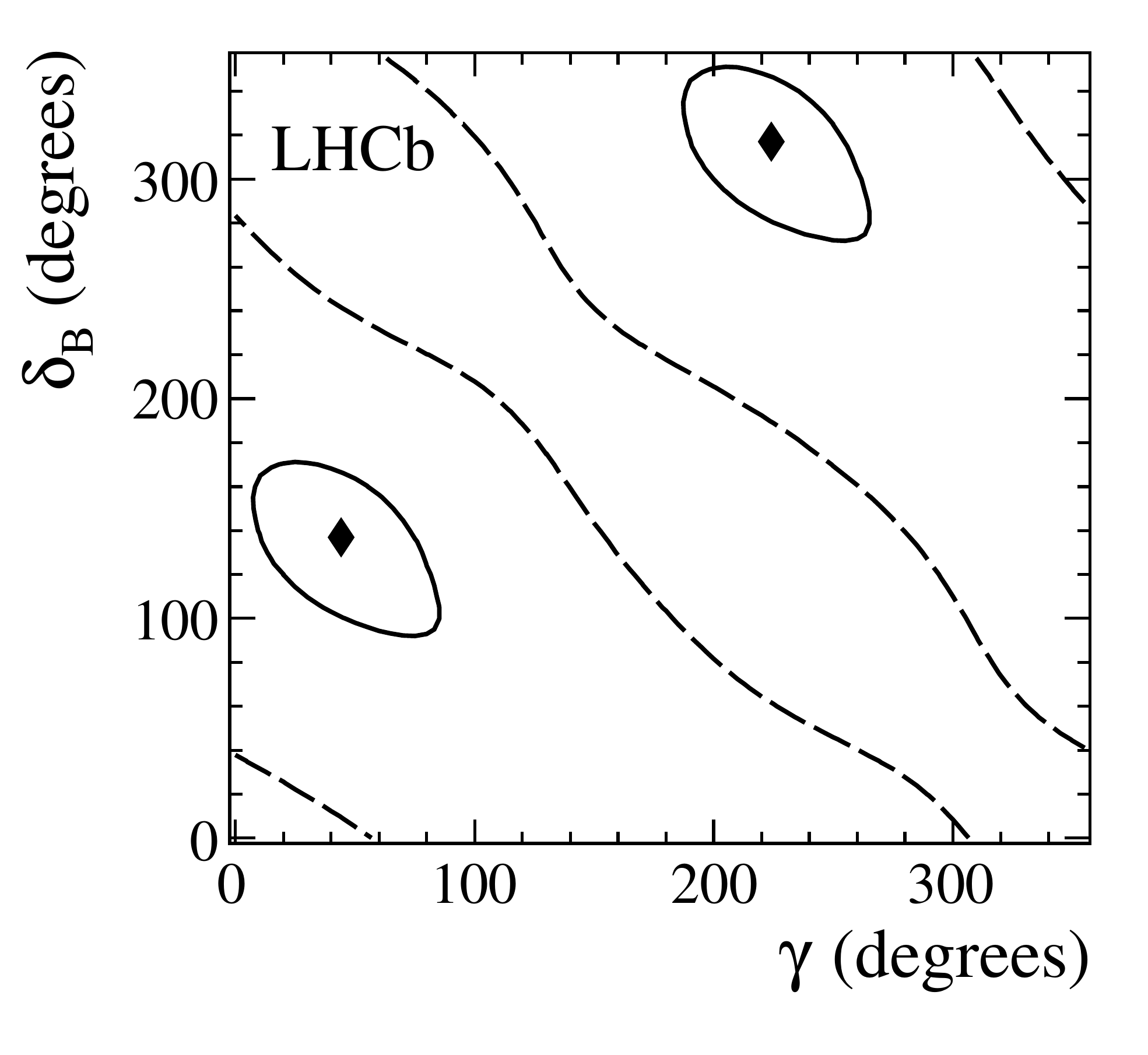}
\caption{\small Two-dimensional projections of confidence regions onto the $(\gamma, r_B)$ and $(\gamma, \delta_B)$ planes showing the one (solid) and two (dashed) standard deviations with all uncertainties included. For the ($\gamma,r_B$) projection the three (dotted) standard deviation contour is also shown. The points mark the central values. }
\label{fig:twodscans}
\end{center}
\end{figure}

The results for $x_\pm$ and $y_\pm$ can be interpreted in terms of the underlying physics parameters $\gamma$, $r_B$ and $\delta_B$.  This is done using a frequentist approach with Feldman-Cousins ordering~\cite{FELDMANCOUSINS}, using the same procedure as described in Ref.~\cite{BELLEMODIND}.  In this manner confidence levels are obtained for the three physics parameters.  The confidence levels for one, two and three standard deviations are taken at 20\%, 74\% and 97\%, which is appropriate for a three-dimensional Gaussian distribution.  The projections of the three-dimensional surfaces bounding the one, two and three standard deviation volumes onto the $(\gamma, r_B)$ and $(\gamma, \delta_B)$ planes are shown in Fig.~\ref{fig:twodscans}. The LHCb-related systematic uncertainties are taken as uncorrelated and correlations of the CLEO-c and statistical uncertainties are taken into account. The statistical and systematic uncertainties on $x$ and $y$ are combined in quadrature.

The solution for the physics parameters has a two-fold ambiguity, $(\gamma, \delta_B)$ and $(\gamma + 180^\circ, \delta_B + 180^\circ)$.  Choosing the solution that satisfies $0 < \gamma < 180^\circ$  yields $r_B = 0.07 \pm 0.04$, $\gamma = (44^{\,+43}_{\,-38})^\circ$ and $\delta_B = (137^{\,+35}_{\,-46})^\circ$.  The value for $r_B$ is consistent with, but lower than, the world average of results from previous experiments~\cite{PDG}.   This low value means that it is not possible to use the results of this analysis, in isolation, to set strong constraints on the values of $\gamma$ and $\delta_B$, as can be seen by the large uncertainties on these parameters.


\section{Conclusions}
\label{sec:conclusions}

Approximately 800 $B^\pm \to D K^\pm$ decay candidates, with the $D$ meson decaying either to $\KS \pi^+\pi^-$ or $\KS K^+ K^-$,  have been selected from 1.0~${\rm fb^{-1}}$ of data collected by LHCb in 2011. These samples have been analysed 
to determine the \CP-violating parameters $x_\pm = r_B \cos (\delta_B \pm \gamma)$ and $y_\pm = r_B \sin (\delta_B \pm \gamma)$,
where $r_B$ is the ratio of the absolute values of the $B^+ \to D^0 K^-$ and $B^+ \to \Dzb K^-$ amplitudes,  $\delta_B$ is the strong-phase difference between them, and $\gamma$ is the angle of the unitarity triangle.   
The analysis is performed in bins of $D$ decay Dalitz space and existing measurements of the CLEO-c experiment are used to
provide input on the $D$ decay strong-phase parameters $(c_i, s_i)$~\cite{CLEOCISI}.   Such an approach allows the analysis to be essentially independent 
of any model-dependent assumptions on the strong phase variation across Dalitz space.  It is the first time this method has been applied to 
$D \to \KS K^+K^-$ decays.   The following results are obtained
\begin{align}
x_-   &=    (0.0 \pm 4.3 \pm 1.5 \pm 0.6) \times 10^{-2}, \; \, &y_-  &= (2.7 \pm 5.2 \pm 0.8 \pm 2.3) \times 10^{-2}, \nonumber\\
x_+   &=    (-10.3 \pm 4.5 \pm 1.8 \pm 1.4) \times 10^{-2}, \; \,  &y_+  &= (-0.9 \pm 3.7 \pm 0.8 \pm 3.0) \times 10^{-2}, \nonumber
\end{align}
where the first uncertainty is statistical, the second is systematic and the third arises from the experimental knowledge of the $(c_i, s_i)$ parameters.
These  values have similar precision to those obtained in a recent binned study by the Belle experiment~\cite{BELLEMODIND}.

When interpreting these results in terms of the underlying physics parameters it is found that  $r_B = 0.07 \pm 0.04$, $\gamma = (44^{\,+43}_{\,-38})^\circ$ and $\delta_B = (137^{\,+35}_{\,- 46})^\circ$.  These values are consistent with the world average of results from previous measurements~\cite{PDG}, although the uncertainties on $\gamma$ and $\delta_B$ are large.  This is partly driven by the relatively low central value that is obtained for the parameter $r_B$. More stringent constraints are expected when these results are combined with other measurements from LHCb which have complementary sensitivity to the same physics parameters.

\section*{Acknowledgements}

\noindent We express our gratitude to our colleagues in the CERN accelerator
departments for the excellent performance of the LHC. We thank the
technical and administrative staff at CERN and at the LHCb institutes,
and acknowledge support from the National Agencies: CAPES, CNPq,
FAPERJ and FINEP (Brazil); CERN; NSFC (China); CNRS/IN2P3 (France);
BMBF, DFG, HGF and MPG (Germany); SFI (Ireland); INFN (Italy); FOM and
NWO (The Netherlands); SCSR (Poland); ANCS (Romania); MinES of Russia and
Rosatom (Russia); MICINN, XuntaGal and GENCAT (Spain); SNSF and SER
(Switzerland); NAS Ukraine (Ukraine); STFC (United Kingdom); NSF
(USA). We also acknowledge the support received from the ERC under FP7
and the Region Auvergne.

\addcontentsline{toc}{section}{References}
\bibliographystyle{LHCb}
\bibliography{main}

\end{document}